\documentclass[%
 reprint,
amsmath,amssymb,
 aps,
pre,
floatfix,
]{revtex4-2}
\usepackage{caption}
\usepackage{subcaption}
\usepackage{epsfig}
\usepackage[utf8]{inputenc}

\usepackage{xcolor}  

\usepackage{graphicx}
\usepackage{dcolumn}
\usepackage{bm}
\usepackage{comment}
\usepackage{multirow}
\usepackage{amsmath}

\begin{document}

\preprint{APS/123-QED}

\title{Characterization of dynamical systems with scanty data\\ using Persistent Homology and Machine Learning} 

\author{Rishab Antosh}
\email{rishabofficial2@gmail.com}

 \author{Sanjit Das}%
\email{sanjit.das@vit.ac.in}
\affiliation{%
Division of Physics,
School of Advanced Sciences,
Vellore Institute of Technology, Chennai Campus,
Chennai, Tamil Nadu – 600 127, India
}%

 \author{%
 N. Nirmal Thyagu}%
 \email{nirmalthyagu@mcc.edu.in}
\affiliation{%
Department of Physics, 
Madras Christian College (Autonomous) [Affiliated to the University of Madras], 
Chennai, Tamil Nadu – 600 059, India
 \\
}%

%


\date{\today}

\begin{abstract}
Determination of the nature of the dynamical state of a system as a function of its parameters is an important problem in the study of dynamical systems. This problem becomes harder in experimental systems where the obtained data is inadequate (low-res) or has missing values. Recent developments in the field of topological data analysis have given a powerful methodology, viz. persistent homology, that is particularly suited for the study of dynamical systems. Earlier studies have mapped the dynamical features with the topological features of some systems. However, these mappings between the dynamical features and the topological features are notional and inadequate for accurate classification on two counts. First, the methodologies employed by the earlier studies heavily relied on human validation and intervention. Second, this mapping done on the chaotic dynamical regime makes little sense because essentially the topological summaries in this regime are too noisy to extract meaningful features from it. In this paper, we employ Machine Learning (ML) assisted methodology to minimize the human intervention and validation of extracting the topological summaries from the dynamical states of systems. Further, we employ a metric that counts in the noisy topological summaries, which are normally discarded, to characterize the state of the dynamical system as periodic or chaotic. This is surprisingly different from the conventional methodologies wherein only the persisting (long-lived) topological features are taken into consideration while the noisy (short-lived) topological features are neglected. We have demonstrated our ML-assisted method on well-known systems such as the Lorentz, Duffing, and Jerk systems. And we expect that our methodology will be of utility in characterizing other dynamical systems including experimental systems that are constrained with limited data. 
\end{abstract}

  
\maketitle


\section{\label{sec:level1}Introduction:}\
Characterization of dynamical systems has been a focal point of research in recent years. In complex systems, parametrical changes can cause the system to bifurcate leading to chaotic phases; these critical parameters are known as bifurcation parameters. However, characterizing the system's dynamics is crucial for identifying optimal performance regions or working regimes. In certain cases \cite{khasawneh2016chatter,sujith2020complex,garfinkel1997quasiperiodicity}, the transition to chaotic regimes can affect the system's performance and efficiency, underscoring the importance of characterizing these transitions. Various methodologies have been developed to describe the dynamics of complex systems from phase space data. Typically, the bifurcation route to chaos is characterized by tracking changes in cycles and subcycles. These transitions result in significant topological changes, making their measurement a worthwhile study.

Persistent Homology (PH) \cite{edelsbrunner2008persistent}, a powerful tool in topological data analysis, can track and analyze the shape of the data. Edelsbrunner et al. \cite{edelsbrunner2002topological} pioneered the work of PH aiming to analyze high dimensional data by extracting topological features. Researchers have extensively applied PH in the context of dynamical systems to derive meaningful insights. For instance,  Perea and Harer \cite{perea2015sliding} attempt to quantify the periodicity of the observed time series by computing PH on the reconstructed state space. PH throws light on the dependence of the persisting features on embedding dimensions and sliding window size. Maletic et al. \cite{maletic2016persistent} employ PH to insist on the importance of the proper choice of embedding dimensions. This was illustrated on systems like Henon Map, Rossler, Lorentz system, and real-world data of ECG and their findings indicated that improper embedding dimension parameters lead to a significant increase in noise within persistence diagrams. Sanderson et al. \cite{sanderson2017computational} applied PH for bifurcation and change-point detection, using a witness complex on reconstructed time series to distinguish different systems. This method is implemented to differentiate the same sound note played on two different instruments and successfully outperforms the conventional classification methods. The work of Myers et al. \cite{myers2019persistent} presents two different approaches to derive a graph construction of the observed time series on which PH is computed to obtain point summaries from the persistence diagrams that can effectively characterize the periodic and chaotic behaviors. Similar work of characterizing the dynamics using PH on reconstructed time series was done in machining dynamics \cite{khasawneh2016chatter} aiming to detect the onset of chatter. Recently, Mittal et. al. \cite{mittal2017topological} focus on characterizing dynamical systems using PH thereby noting the transition towards chaos. They make use of relevant $k-D$ features to detect the subcycles and the chaotic behavior. 

Although these methodologies mentioned above use PH on point cloud data or reconstructed phase space data to characterize the dynamics of the system, the characterization of sparse or scanty phase space data can be a significant and interesting study. Specifically, this challenge is critical in situations where there is no prior information about the nature of the systems and the phase space data obtained is incomplete, as often encountered in real-world scenarios. We say this because real-world and experimental data are prone to missing elements or measurements, leading to sparse or scanty datasets \cite{pavlova2019effects}. However, one might be tempted to experiment with the traditional model-based \cite{sauer1998spurious,rosenstein1993practical,wolf1985determining} time series analysis by measuring any one of the state vectors (considering it as a time series data). These traditional approaches, however, require the time series to either have the governing equations or the time series to be sequentially sampled (with no missing elements) and sufficiently long. Given these limitations, we instead adopt a data-driven approach in the form of PH to analyze the available scanty phase space data. PH allows us to compute barcodes \cite{ghrist2008barcodes}, which capture both long-lived and short-lived topological features of the data. However, distinguishing between true features and noise within these barcodes is challenging due to the absence of a clear threshold for classification; this process typically requires human visual inspection. Therefore, there is a need for a more advanced tool to effectively automate this classification. To address this, we employ a machine learning (ML) algorithm, specifically a binary classifier, to differentiate true features and noise in the barcodes.

The combining ideas of implementing PH and ML together on sparse or scanty phase space data led us to interesting insights about the system with which we devise topological summaries. We propose two topological summaries: Persistence score and Noise score that help us to characterize the transitioning region from periodic to chaotic. We practice our proposed methodology on two-dimensional and three-dimensional complex systems to showcase their versatility.

\section{Background on Homology and Persistent Homology}

Interpretation of data has been the intense focus of study for the past few decades. Topological Data Analysis (TDA), a relatively new field of research has provided breakthroughs in many areas including the field of medicine \cite{nicolau2011topology,crawford2016topological,emmett2014parametric}, robotics \cite{bhattacharya2015persistent,vasudevan2013persistent}, and dynamical systems\cite{maletic2016persistent,myers2019persistent,khasawneh2016chatter}. TDA combines algebraic topology, statistics, data analysis, and computational topology under the same roof to provide a meaningful qualitative analysis of the dataset under study. Persistent Homology (PH) \cite{edelsbrunner2002topological,zomorodian2001computing}, a powerful tool in the branch of TDA, is widely used in many places and it is mainly known for its robust nature, stability to noise, and easy interpretability of topological features obtained. PH can be used to analyze finite metric spaces, real-valued functions, and digital images.

\subsection{Simplicial Complex:}
We define a $p$-simplex $\sigma^p$ as: For an affinely independent set of $(p + 1)$ points $(v_0, v_1, v_2, \ldots, v_p \in \mathbb{R}^p)$, a $p$-simplex is the set of points. 
\[
\sum_{i=0}^{p} \lambda_i v_i
\]
such that
\[
\sum_{i=0}^{p} \lambda_i = 1 \quad \text{where} \quad \lambda_i \geq 0 \quad \text{for all } i.
\]
By this, we mean that every point in the $p$-simplex $\sigma^p$ is an affine combination of the points $(v_0, v_1, v_2, \ldots, v_p)$, and the coefficients $\lambda_i$ determine the positions of the points in the simplex. Hence, a simplex of dimension $p$ in an Euclidean space is the convex hull of $p+1$ affinely independent points. In general, a simplex is a generalized triangle of any dimension where $0$-simplex represents a vertex, $1$-simplex represents an edge, $2$-simplex represents a triangle, and so on. Hence we go on by defining that a simplex of $p$ dimension is the smallest convex shape that can be built by joining $p+1$ points (or vertices) through edges. 

Simplicial complex K is a collection of simplices of different dimensions glued together as a single unit. The intersection of any two simplices must be either empty or must have a common face, this makes sure that they do not overlap. By this, we mean that the non-empty intersection of any two simplices $\sigma_1, \sigma_2$                     
is a face of both $\sigma_1$ and $\sigma_2$. The dimension of the simplicial complex K is the highest dimension of the simplices that it includes.

\subsection{Chain complex and Homology:}

A $p$-chain $c \in C_p$ is a linear combination of $p$-simplices in a simplicial complex $K$. In other words, $C_p$ is a collection of p-chains formed by taking linear combinations of simplices, and it forms a group structure where the addition and subtraction of $p$-chains are defined. We can denote this as: $c = \sum_{i} a_i\sigma_i$ where $a_i$ represents the coefficients and $\sigma_i$ represents the  
$k$-simplicies. Two $k$-chains are represented as: $c = \sum a_i\sigma_i$ and $c^{'} = \sum b_i\sigma_i$ then, the addition operation is defined as $c + c^{'} = \sum (a_i + b_i)\sigma_i$ and we take the coefficient group as ${\mathbb{Z}}_2$ (modulo 2) which implies $1+1=0$. To relate or map the chain groups, we define a boundary operator and it is denoted by $\partial$. For a simplex $\sigma = [v_0, v_1, \ldots, v_k]$ its boundary is defined as:
\[
\partial_p \sigma = \sum_i (-1)^i [v_0, v_1, \ldots, \hat{v_i}, \ldots, v_p],
\]
where $\hat{v_i}$ denotes the omitted vertex. The boundary $\partial_p(\sigma)$ of a $p$-simplex $\sigma$, maps $p$-chains from $p$ to $p-1$ dimension or in other words, the boundary operator maps $\partial_p : C_p \to C_{p-1}$. The boundary of a $p$-chain is the sum of the boundaries of its simplices: $\partial_p c = \sum a_{i} \partial_p \sigma_i$. Now, it is understood that each boundary operator is a homomorphism $\partial_p: C_p \rightarrow C_{p-1}$, and they map the chain groups into a chain complex as:
\[
\cdots \xrightarrow{\partial_3} C_2 \xrightarrow{\partial_2} C_1 \xrightarrow{\partial_1} C_0 \xrightarrow{\partial_0} 0
\]

The property that the boundary of a boundary is zero implies that the image of the $(p + 1)$-th boundary map must be a subset of the kernel of the $p$-th boundary map:
$\partial_p \circ \partial_{p+1} = 0$ and hence, $\text{Im}(\partial_{p+1}) \subset \text{Ker}(\partial_p)$. So we write that elements of $B_p := \text{Im}(\partial_{p+1})$ are referred to as boundaries and the elements of $Z_p := \text{Ker}(\partial_p)$ are called cycles.

Now the $p^{th}$ Homology group ($H_p$) of a simplicial complex K can be defined as a quotient group:  \[H_p(K) = Z_p/B_p
\] 
In simple terms, $p^{th}$ homology group detects the p-dimensional holes where, $H_0(K)$ detects the 0-dimensional holes (which are essentially the connected components), $H_1(K)$ detects the 1-dimensional holes (loops or circular holes), $H_2(K)$ detects the 2-dimensional holes ( voids or tunnels) and so on. The $p^{th}$ Betti number $\beta_p$ is defined as the rank of the homology group $H_p(K)$         \[\beta_p = \text{rank}(H_p(K))
\]
$\beta_p$ essentially counts the number of $p$-dimensional holes in $K$.

\begin{figure*}
\includegraphics[width=0.7\textwidth]{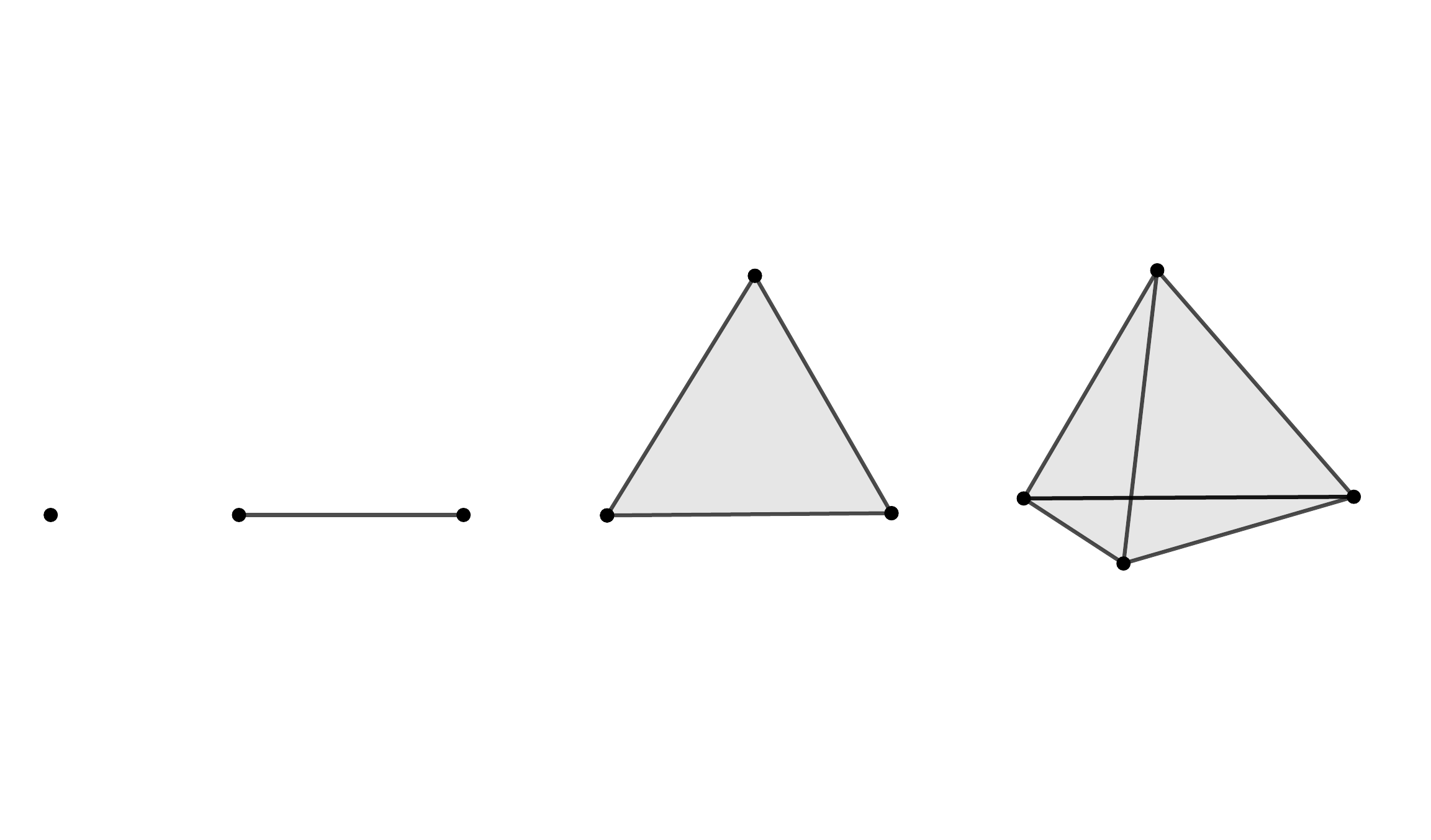}
    \caption{Simplex representation in different dimensions, 0-simplex, 1-simplex, 2-simplex and 3-simplex from left to right respectively.}
    \label{fig:simplex combo}
\end{figure*}

\begin{figure}
\centering
\includegraphics[width=6.8cm,height=4.0cm]{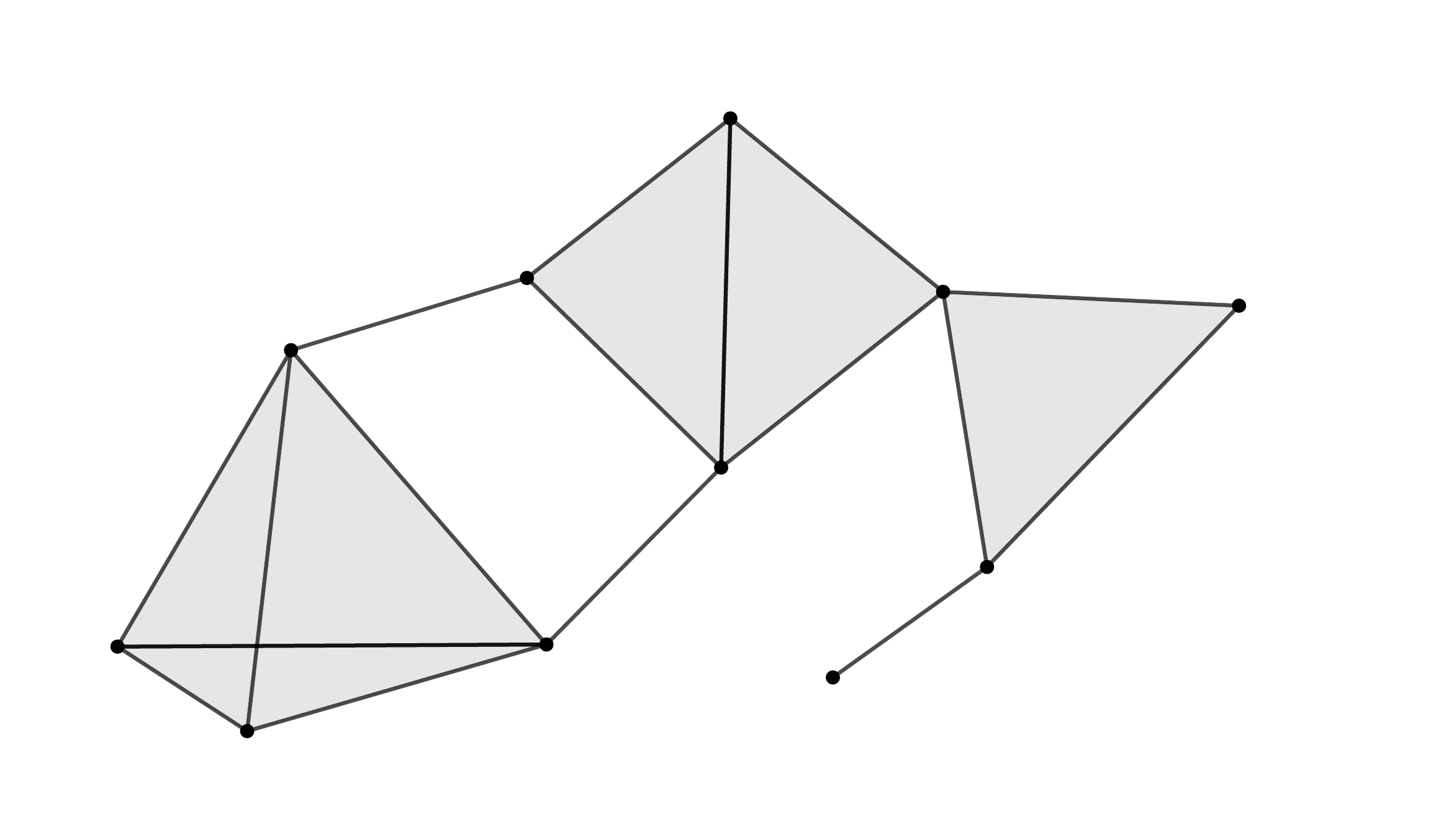}
\caption{Simplicial complex with a 3-simplex, 2-simplex, and 1-simplex combining to form a single unit. This visual representation also shows how two simplices combine to share a common face.}
\end{figure}

\subsection{Filtration process:}
Persistent Homology (PH) is a method that allows the sequential growth of a simplicial complex to track the topological structure of the data through a process called filtration. It quantifies the persistence of the topological features as it undergoes the filtration process. There are quite a few filtration processes available to build this nested sequence of subcomplexes, in the form of Cech complex filtration, Vietoris-Rips Complex filtration, alpha complex filtration, and so on. We have chosen Vietoris-Rips Complex filtration [13] because of its computational efficiency and flexibility (requires pairwise distances). In short, given point cloud data, PH extracts the shape of the data. We construct the Vietoris-Rips Complex for a given set of points using Euclidean distance as a filtration radius ($\epsilon$) to build the V-R complex ($V_\epsilon$). This $V_\epsilon$ contains all simplices whose radius is less than or equal to the threshold parameter $\epsilon$. The general idea is to draw Euclidean balls of radius $\epsilon$ and build $p$-simplex for each $p+1$ point whose pairwise distances are less than $\epsilon$. V-R complexes are added one after the other for increasing $\epsilon$ which will give rise to filtration, provided the simplicial complexes ($K_n$) will be a subset of its successor ($K_{n+1}$) (for increased values of $\epsilon$). The $P_{th}$ homology group $H_p$ is defined by removing cycles from the higher dimensions, in short, this calculates all the boundaries and removes the boundaries coming from the higher dimensional objects, and the leftover is counted. During the filtration process, when a $p$-dimensional hole is recorded, we have $\epsilon_1$ and $\epsilon_2$ coordinates noted which corresponds to the birth and the death of that feature. The Betti number ($\beta_p$) keeps track of the number of $p$-dimensional holes which can be visualized in two ways: Persistence Diagram (PD) and barcodes. PD is a plot representing the birth and death coordinates (filtration parameter $\epsilon$) of features arising for various homological groups. Generally, we look for the points that are far away from the diagonal and assert that as long-lived or the true features while the points closer to the diagonal are considered to be noise. PDs are often interpreted in terms of barcodes to show the lifetimes (or longevity) of these features. The lifetime of a feature is computed by taking the difference between the death and birth coordinates of the features in PD. In a barcode plot, long-lived features have extended lifetimes, while short-lived features have brief lifetimes.

\begin{figure*}
\includegraphics[width=0.8\textwidth]{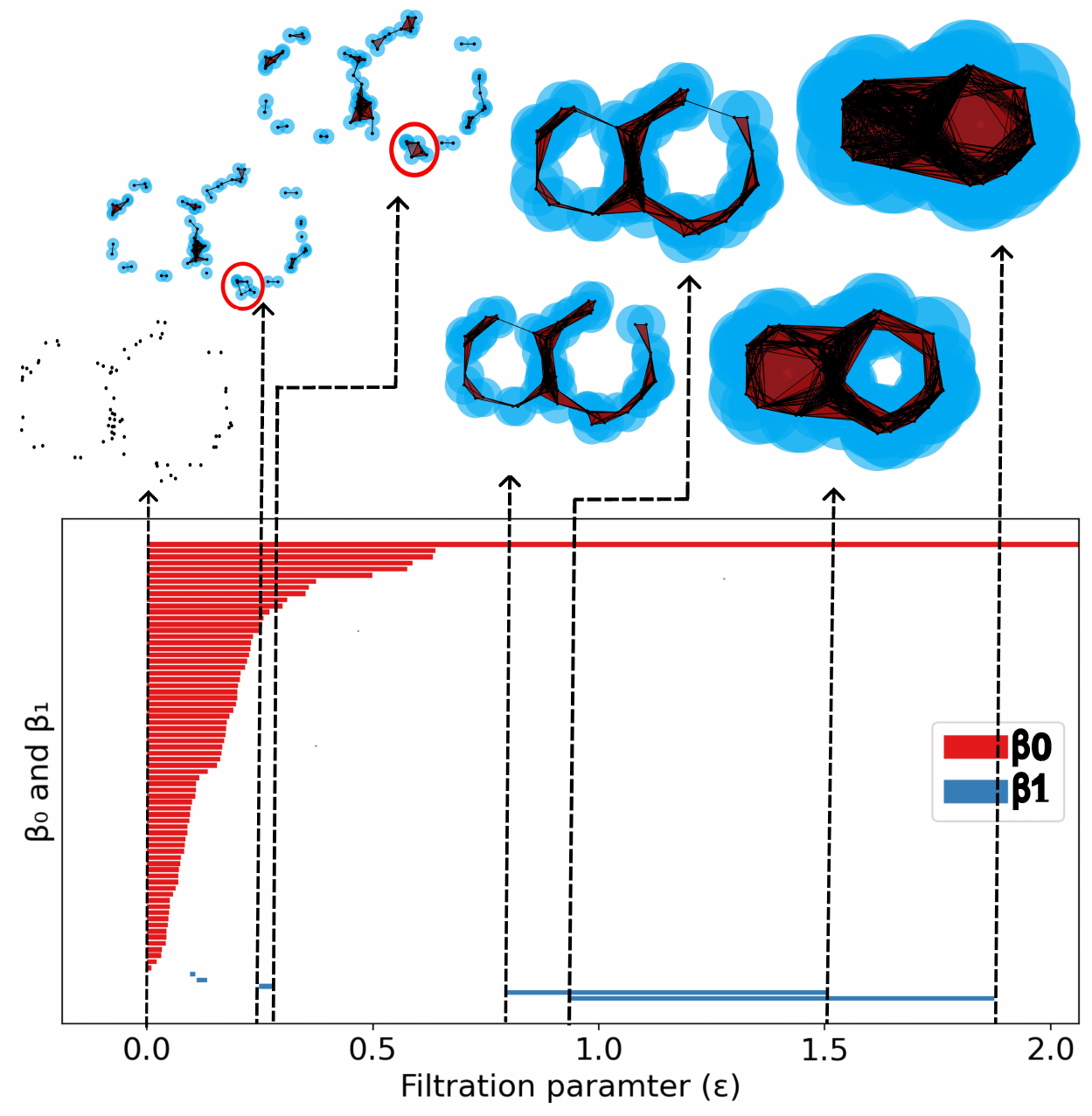}

    \caption{This figure illustrates the filtration process of the synthetic point cloud data (PCD) mimicking figure `8': The red barcodes denote the connected components ($\beta_0$) and the blue barcodes denote the holes ($\beta_0$). The focus is on the birth and death of 1-D holes: The black dotted lines provide a one-to-one correspondence between the barcodes and the V-R filtration \footnote{We use Mathematica software \cite{vrcechmathematica} to visualize the V-R construction process} in PCD, demonstrating how barcodes \footnote{We use Gudhi library \cite{gudhi:RipsComplex} to visualize the sequential arrangement of barcodes for varying filtration paramter} capture the creation and annihilation of holes ($\beta_1$). The PCD exhibits two major enclosures, analogous to Figure `8', the barcodes validate the topological structure by producing two persisting blue barcodes termed features while the smaller blue barcodes are termed noise.}
    \label{fig:PH_filtration}i
\end{figure*}

Figure \ref{fig:PH_filtration} represents the filtration process of a synthetic point cloud data which has the shape of `8' for demonstration. The longevity of the $\beta_0$ and $\beta_1$ features are represented as barcodes. Here, $\beta_0$ (red lines) represents the number of connected components and $\beta_1$ (blue lines) represents the number of created $1-D$ holes. We focus only on $\beta_1$ features which are essentially more important than the $\beta_0$ in analyzing the shape and structure of the point cloud data. The illustration aims to show the emergence and destruction of long-lived (prominent) and short-lived features and their interpretation in terms of barcodes. We provide a one-to-one correspondence between the barcodes and the V-R filtration process using the black dashed lines to show the emergence and death of the $H_1$ features. The second dashed black line shows the emergence of a transient hole (connotated as noise) at $\epsilon=0.248$ and the third dashed line represents the death (closure) of the transient hole (noise) by filling the hole using $2-simplex$ and $3-simplex$ (highlighted in red circle) for the increasing filtration parameter of $\epsilon=0.278$. The fourth and fifth dashed black lines represent the emergence of two of the long-lived hole (feature) at $\epsilon = 0.8$ and $\epsilon = 0.935$ respectively while the sixth and seventh dashed black line represents the death of these two long lived holes at $\epsilon = 1.51$ and $\epsilon = 1.88$ respectively. In short, The point cloud data representing the shape of `8' essentially has two prominent loops. We see two significant $\beta_1$ barcodes capturing these loops in the barcode diagram.  In addition to these, there are other short-lived $\beta_1$ barcodes present, which are generally discarded as noise. These short-lived barcodes arise because of the transient holes that appear during the filtration process.

\begin{figure}[h!]
     \centering
     \begin{subfigure}{0.31\textwidth}
         \centering
         \includegraphics[width=\textwidth]{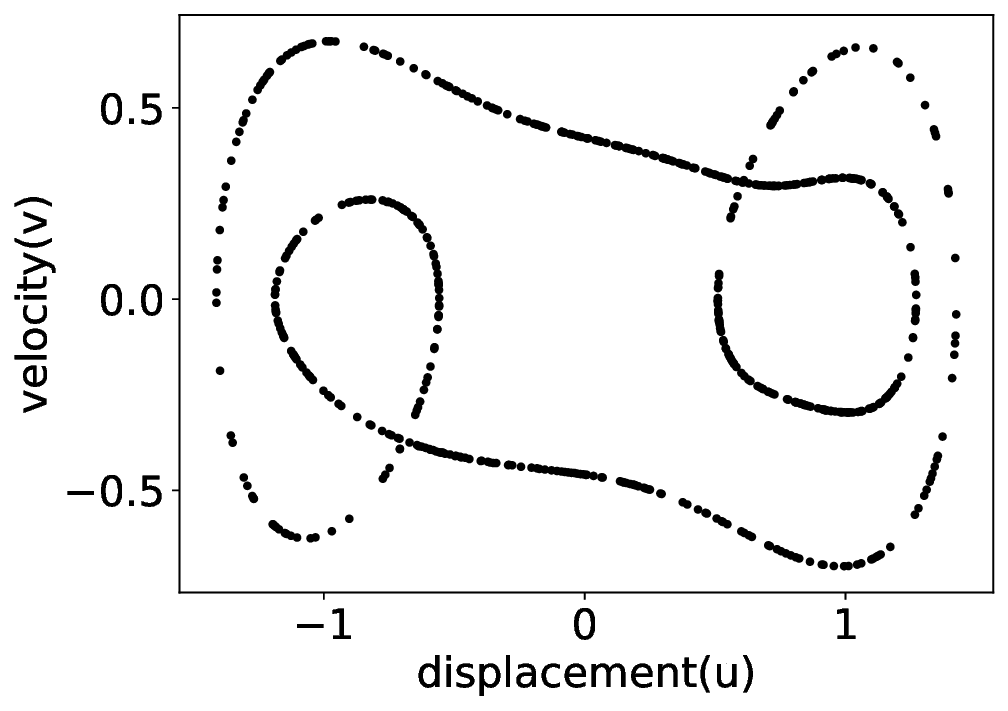}
         \caption{Phase space data for period-3 duffing system with 600 landmarks.}
         \label{duff_PCD}
     \end{subfigure}
    \hfill
     \begin{subfigure}{0.31\textwidth}
         \centering
         \includegraphics[width=\textwidth]{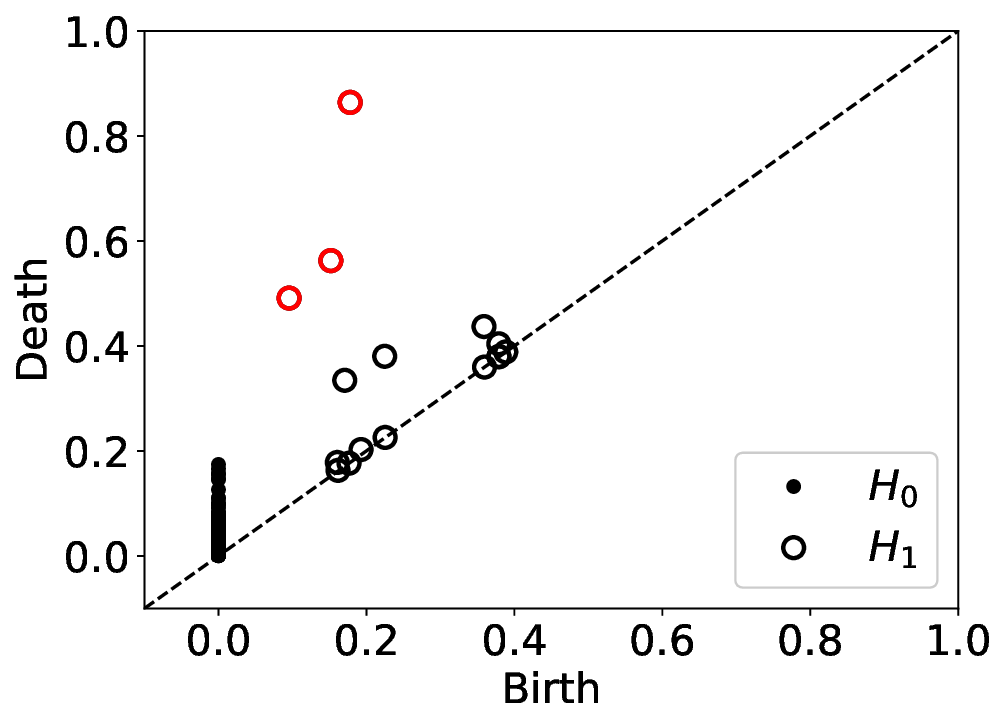}
         \caption{Represent the Persistent diagrams with $H_0$ and $H_1$. We have three persisting features or holes (red circles) in $H_1$ that are far away from the diagonal.}
         \label{duff_PD}
     \end{subfigure}
     \hfill
     \begin{subfigure}{0.31\textwidth}
         \centering
         \includegraphics[width=\textwidth]{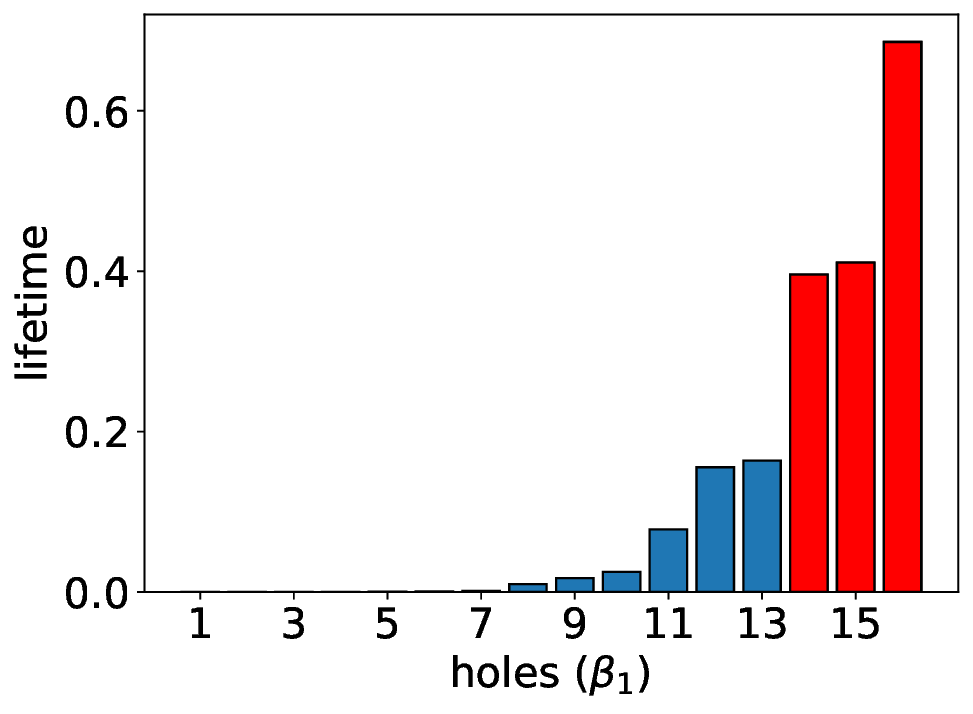}
         \caption{Represents the $\beta_1$ barcodes: lifetimes of $H_1$ holes. The three long barcodes (red barcodes) represent the period-3 nature of the system.}
         \label{duff_barcode}
     \end{subfigure}
              \caption{This plot represents the computation of PH for a phase space data of a duffing system exhibiting period-3 behavior. The Persistence diagram (PD) shows the existence of two homological groups: $H_0$ and $H_1$. We are interested in analyzing the system's dynamical state, which is generally done by tracking changes in the subcycles. Hence, we study only $\beta_1$ which counts the number of holes. We have three long-lived features $H_1$ that are highlighted in red circles. Alternatively, we have the barcode representation which denotes the longevity (or lifetimes) of holes that appear. Barcodes indicating the three long-lived features are highlighted in red color hinting that it is a period-3 system.} 
        		\label{fig:duff_filtration}
     \end{figure}

We extend the computation of PH in dynamical systems to illustrate the importance of detecting their topological features. Consider Fig. \ref{fig:duff_filtration}, which illustrates the PH computation for phase space data comprising 600 landmarks derived from the Duffing system. Visual inspection of the phase space data (Fig. \ref{duff_PCD}) reveals that the system exhibits a period-3 behavior, as indicated by the subcycles present in the data. Computation of PH for the phase space data yields PD that represents two homological groups where the black dots denote the number of connected components ($H_0$) and black circles denote the number of 1-D holes ($H_1$). Fig. \ref{duff_barcode} represents the barcodes arising from $H_1$, we can see three prominent barcodes that denote the three subcycles in the point cloud data. The three long-lived (prominent) barcodes are regarded as the actual topological features while the other 13 short-lived barcodes of $H_1$ are considered as noise. In dynamical systems, bifurcations are characterized by qualitative changes in the system's behavior as parameters are varied. These changes are often reflected in the subcycles' changing numbers and sizes, insisting on the importance of tracking their evolution. In the subsequent analysis of this study, we focus on $\beta_1$, as it tracks the morphology of 1-D holes or loops, providing insights into the dynamical state of the system. Conversely, the $\beta_0$ study, which essentially counts the number of connected components is set aside.

In this paper, the PH computations were performed using the Ripser library from scikit-tda \cite{scikittda2019} in Python. The computations were run on a computer with an Intel(R) Core(TM) i5-8300H processor with a CPU speed of 2.30 GHz using 8.00 GB RAM running on Windows 11. It took a maximum average time of 3 seconds to compute PH for one PCD of 600 landmarks.

However, validating these long-lived barcodes as true features requires human eye validation by comparing the barcode data with the system’s phase space data, establishing a one-to-one correspondence between the long-lived features and the subcycles (we look at the number of loops to determine its periodicity) of
the system. 
\section{Machine learning classifier:}
One of the main objectives of using persistent homology in the study of dynamical systems is to determine the dynamical state of a system at a given value of the system parameter(s). However, in the case of Fig. \ref{fig:duff_filtration}, the three barcodes were acknowledged as features primarily based on one-to-one correspondence to the phase space data (we look at the number of loops in the phase space data and look at the perceived long-lived barcodes). If one merely were to look at the barcode data, one could also pick only the last barcode (the 16th) or the last six barcodes (11th to 16th) as true features without looking at the phase space data. Hence, the visual inspection approach has two limitations. First, there is no clear threshold to distinguish between noise and feature. A threshold value for a barcode is necessary to tell us how long a barcode must persist to be accepted as a true feature. Second, it is a laborious process; manual counting of large sets of phase space datasets obtained for a range of parameters to identify and count cycles or sub-cycles with the extracted barcodes is time-consuming and it is prone to human inaccuracies. Alternatively, one might be tempted to employ any simple statistic like the mean, standard deviation, etc to extract a handy threshold value for discriminating significant barcodes from the insignificant barcodes. But one quickly realizes that no such easy solution exists. For instance, if we use the mean value of the barcode lengths to get a threshold length scale, we observed that instead of three barcodes as significant features the criterion benchmarked five barcodes. Likewise, any other attempt to arrive at the threshold value of barcode lengths using simple statistics will not lead to any improvement because of the need to validate the barcodes with the corresponding phase space sub-cycle information. All these statistical methods rely only on the distribution of the data without weighting important features resulting in a less accurate threshold determination. So, it is suggestive that visual inspection of the barcodes or simple statistical calculations as classification threshold would fail without the apriori visual information of the phase space. This motivated us to introduce a machine learning (ML) based binary classifier in the paper. We use a supervised ML model, where the input data is labeled. ML learns the patterns from the labeled data, separates them into two categories (binary), and computes a threshold to classify the unseen data into its respective categories. But the advantage of using ML is that it uses a weighted linear combination of the input features and employs gradient descent which iteratively updates weights based on the prediction error or loss function (the difference between the predicted outcomes and the actual labels) till it is fully optimized. 
This addresses two key issues: One, it computes the threshold for the classification of features and noise based on the data fed; Second, it is an automated process once the model is trained. Binary classifiers fit in when target data are categorical, in our case it is distinguishing true features and noise, hence we can consider the feature barcode as 1 and the noise barcode as 0. We use the Logistic regression (LR) model as the binary classifier for this problem.

Logistic regression in ML is one of the most important methods in binary classification problems \cite{friedman2001elements,agresti2018introduction}. The advantage of LR is it can be extended to multi-class classification and the output values are probabilistic \cite{friedman2001elements} \cite{multiclass}. LR is used in various fields, in the medical arena to infer the severity of the disease and by giving a score to indicate \cite{article,BIONDO2000635} and can used for early detection of developing a chronic disease by assessing various parameters from the medical tests \cite{truett1967multivariate}, and used in other fields like politics \cite{harrell2017regression} and engineering \cite{PALEI200988}. The sigmoid or logistic function is an S-shaped curve that maps a real-valued number to a range between 0 and 1. Now, the sigmoid function is defined as \begin{math}\eta(Z)=\frac{1}{1+e^{-z}}\end{math}, where the regression function is given as input to the sigmoid or logit function. So in a nutshell, LR is a prediction of the probability of occurrence of an event by fitting a logistic curve.

\subsection{Generation of the input data for the machine learning classifier:} \label{sec:input generation}
We choose random 600 landmarks as sampled phase space data for all the dynamical systems to recreate real-world data. Computation of PH on numerically evolved phase space data can be computationally intensive. For constructing a Vietoris-Rips (V-R) complex from a point cloud data containing N points, the number of simplices grows exponentially (approximately $2^N$). To counter this complexity, practitioners generally downsize (sample) the point cloud data using minimax sampling or a coarse-graining procedure. These sequential sampling methods aim to provide a representative subset of data in such a way that it retains the global and local structure of the original data. However, our main intention in randomly sampling the data is to mimic the experimental data with missing elements, essentially losing the information of the phase space structure. However, this random sampling approach not only helps in recreating real-world scenarios but also reduces the computational expense as a latent benefit.

In this paper, we use a simple approach, wherein we utilize just one phase space data with 600 landmarks (we take $\gamma = 0.35$ in the duffing system) as a given input. We know apriori that at this parameter value of $\gamma$ the system is in the periodic regime. Our task is to use the only data set given to us (at $\gamma = 0.35$) to prepare the ML algorithm for it to classify the state of the system at other parameter values of $\gamma$. The advantage of this approach is that for determining the phase transitions, we do not demand more than one input. Here, our input data needs to be from a periodic state of the phase space. The disadvantage of one data set, as any ML practitioner would point out, is the lack of sufficient data to train the ML algorithm. To offset this problem, we use the 600 landmarks data and generate a number of subsets of this data by randomly extracting points.  The obtained subsets contain the following number of landmarks: 575, 550, 500, 475, 450, and 300 (to compute PH and extract the barcodes). Although these data sets originate from a single value of the parameter $\gamma$, we have sampled them randomly to acquire a collection of datasets that appear to have come from multiple instances. The number of subsampled data and its size may vary depending on the specific data, but generally, sufficient sampling is performed until optimal evaluation results are achieved. Now we have a considerable amount of data that can be used to train the model and test the ML algorithm to evaluate the performance of the model.

Consequently, this approach can introduce a minor drawback, as we shall see shortly, due to the consideration of sparser landmark data. Such sparser data generate barcodes that are slightly less long-lived because as the landmark size decreases, the features' longevity also reduces. This can have an impact on the classification threshold which is computed as a result. Alternatively, we also have another approach that counters the hurdle mentioned above: We can use a range of periodic regime bifurcation parameters (from $\gamma=0.35$ to $\gamma=0.38$) as the datasets as given inputs. This method is seen to give a better classification threshold since the inputs come from the $\gamma$ parameters of similar landmark size, 600. But, we stick to the first approach considering the minimalistic requirement of the data which often happens in real-world scenarios, but one can always opt for the second approach considering the availability or measurement of data for other periodic parameters if possible.

The obtained barcodes are manually (human-assisted classification) classified as feature and noise based on the subcycles (periodicity) of the system. The LR classifier model in ML divides the input dataset into two sets: train and test set.  Now, using the train set the LR model learns the patterns based on the labeling and computes a threshold value for classification. Now using this trained model it classifies the test data and evaluates the performance of the model.

\subsection{Evaluation metrics:}
We evaluate our model based on the test set split and this evaluation result can serve as a litmus test for the selection of input data. Evaluation metrics in ML are an essential entity as it assesses the performance of the model. Quantifying the performance of the model is important as it can hamper the prediction of new target variables leading to undesirable results. Evaluation metrics can also act as a litmus indicator for the proper choice of the training set paving the way to make the necessary changes. We make use of some commonly used metrics to showcase the performance of the model, and the metrics are mentioned below:
\begin{enumerate}

\item Confusion matrix: It is a $N \times N$ matrix \cite{raschka2019python}, where N is the number of classes or categories. In our case, it is a binary classifier problem (0 or 1), hence N=2. It is a matrix of four components: true negatives (TN) which represent the case where the predicted category is 0 and the actual category is 0, false positives (FP) which represent the case where the predicted category is 1 while the actual category is 0, false negatives (FN) which represent the case where the predicted category is 0 while the actual category is 1, true positives (TP) which represent the case where the predicted category is 1 and the actual category is 1. 

\item Accuracy: It is defined as the ratio of the number of correct predictions to the total number of samples (total data) \cite{raschka2019python}. This gives us information about the correctly classified samples. This metric is important because too many misclassified samples will change the destiny of the nature of the system. Mathematically accuracy can be represented as: 
\\ Accuracy = $\frac{TP + TN}{TP + FN + TN + FP}$

\end{enumerate}

This paper computed the ML computations using the scikit-learn \cite{sklearn_api} in Python. The computations were run on a computer with an Intel(R) Core(TM) i5-8300H processor with a CPU speed of 2.30GHz using 8.00 GB RAM running on Windows 11.

\section{Topological summaries:}
\begin{figure}[h!]
     \centering
     \begin{subfigure}{0.45\textwidth}
         \centering
         \includegraphics[width=\textwidth]{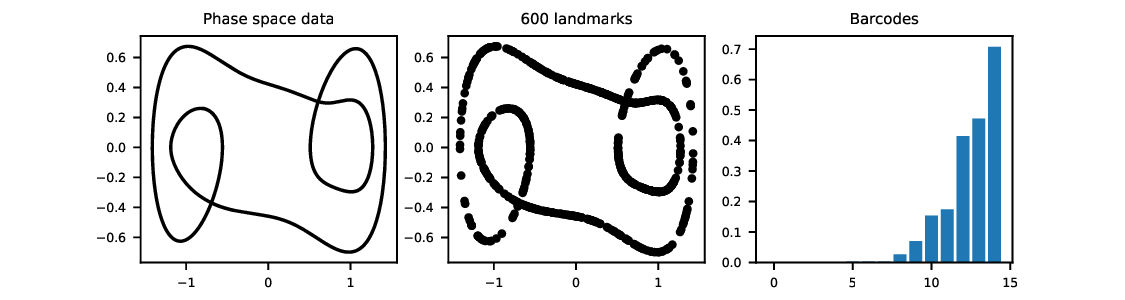}
         \caption{Barcode of $\gamma = 0.35$ (period-3) for 600 landmarks.}
         \label{fig:0.37_600}
     \end{subfigure}
     \begin{subfigure}{0.45\textwidth}
         \centering
         \includegraphics[width=\textwidth]{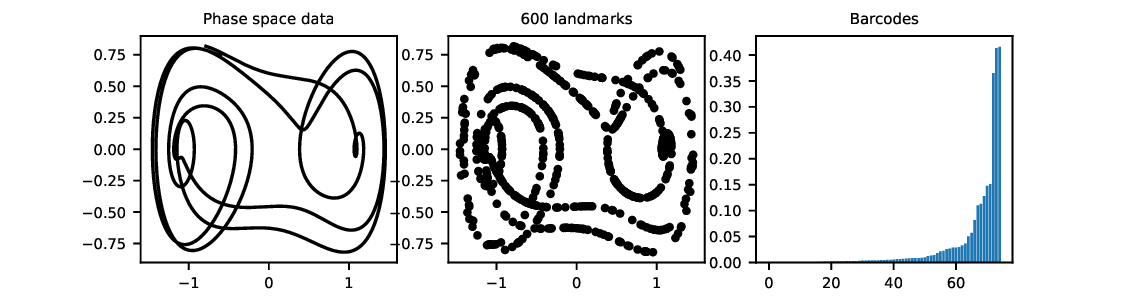}
         \caption{Barcode of $\gamma = 0.39$ (chaotic) for 600 landmarks.}
         \label{fig:0.39_600}
     \end{subfigure}
     	        \caption{Comparison of barcodes obtained by computing PH for $\gamma = 0.35$ (period-3) and $\gamma = 0.39$ (chaotic) for 600 landmarks of duffing oscillator respectively: The chaotic phase portrait exhibits irregular and complex trajectories that contribute to the increased number of noise barcodes than compared to the periodic regime.}
        		\label{fig:noise comparison}
     \end{figure}

\begin{figure*}
\centering
\begin{subfigure}[b]{1\textwidth}
\includegraphics[width=1\textwidth]{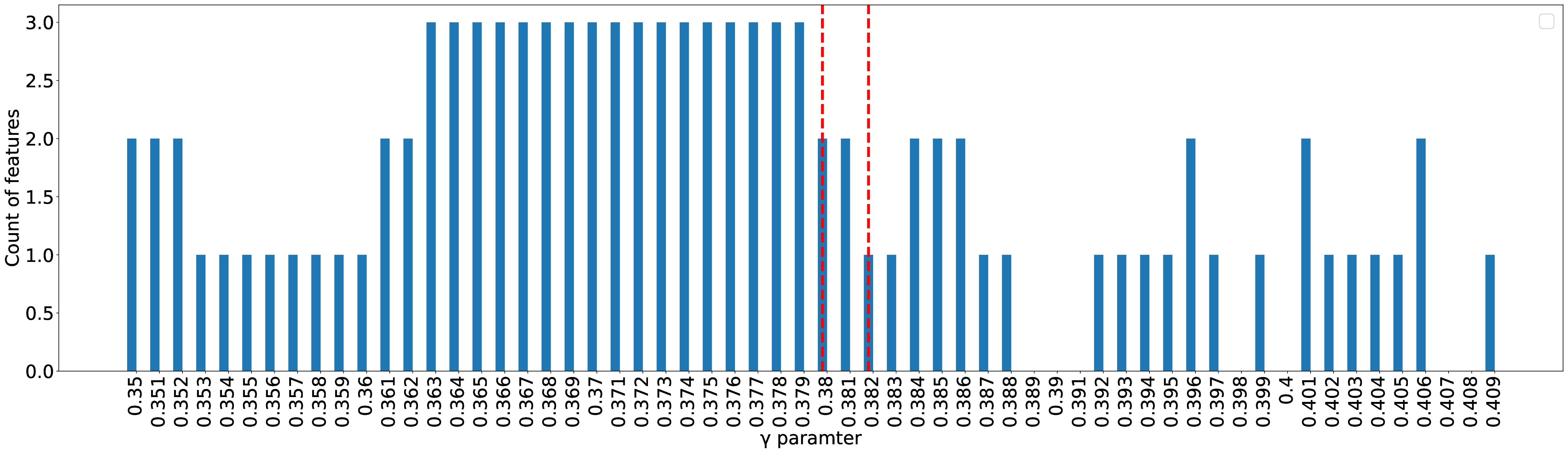}
\caption{}
\label{fig:barplot_features}
\end{subfigure}
\begin{subfigure}[b]{1\textwidth}
\includegraphics[width=1\textwidth]{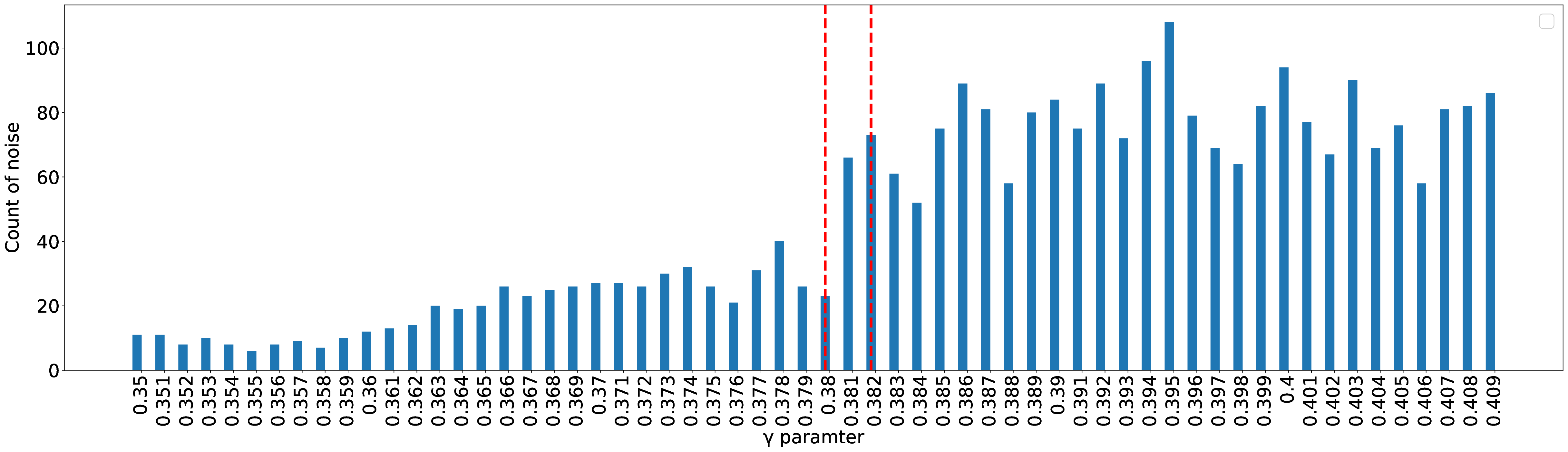}
\caption{}
\label{fig:barplot_noise}
\end{subfigure}
\caption{
These plots present the counts of (a) features and (b) noise as a function of the parameter $\gamma$ that was classified by our ML algorithm (\emph{Scheme 1}). 
The ML algorithm in \emph{scheme 1} was trained with barcodes from a single periodic regime phase space data that was further subsampled into multiple data sets with 600, 575, 550, 500, 475, 450, 400, 350, 300 landmarks at $\gamma = 0.35$. The red dotted lines on both plots represent the transition region from the periodic to the chaotic regime. We report that counting the classified features seems trivial since no significant change is observed that can actively characterize the periodic and the chaotic regime transition. On the contrary, we observe that counting the classified noise seems to be a good quantifier since we could see a significant change in the numbers as we traverse into the chaotic regime.}
\label{fig:barplot comparison}
\end{figure*}

\begin{figure*}
\centering
\begin{subfigure}[b]{1\textwidth}
\includegraphics[width=1\textwidth]{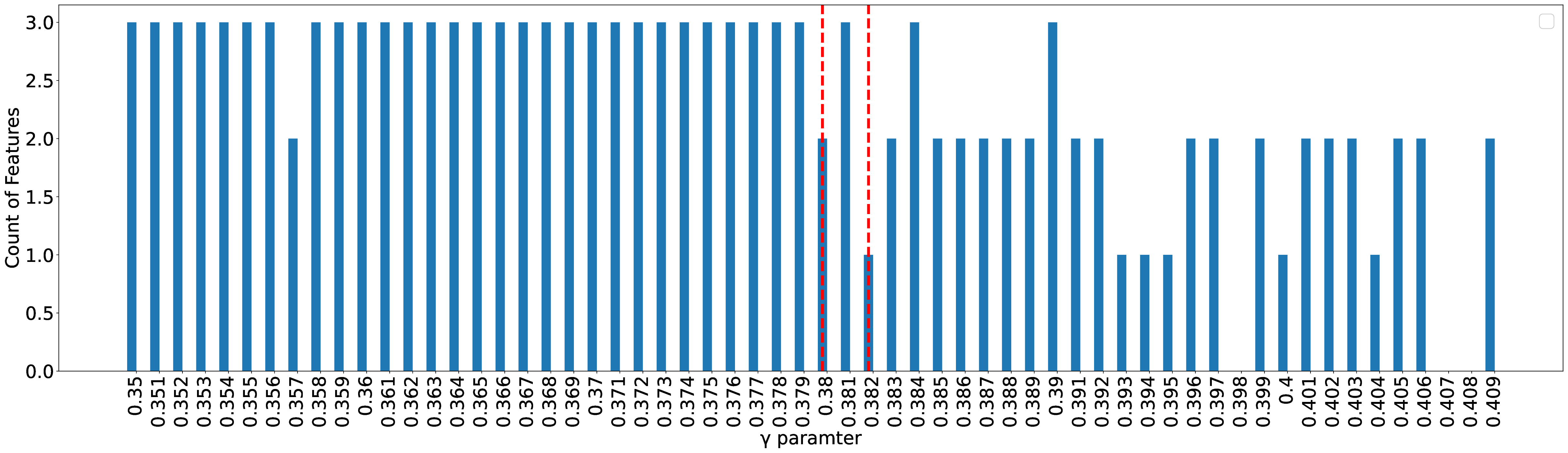}
\caption{}
\label{fig:barplot_range_F}
\end{subfigure}
\begin{subfigure}[b]{1\textwidth}
\includegraphics[width=1\textwidth]{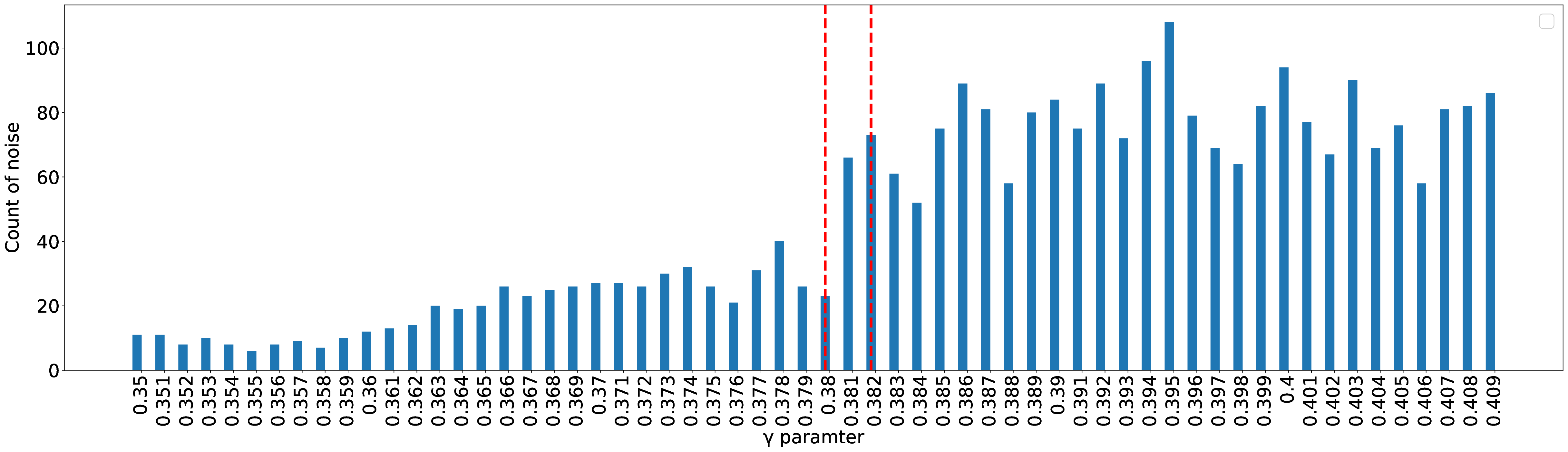}
\caption{}
\label{fig:barplot_range_N}
\end{subfigure}
\caption{These plots present the counts of (a) features and (b) noise as a function of the parameter $\gamma$ that was classified by our ML algorithm (\emph{Scheme 2}). The ML algorithm in \emph{scheme 2} was trained with barcodes from the periodic regime phase space dataset in the range $\gamma = 0.35$ to $\gamma = 0.38$. The red dotted lines on both plots represent the transition region from the periodic to the chaotic regime. unlike Fig. \ref{fig:barplot_features}, we have a consistent count in the periodic regime and fluctuating counts in the chaotic regime but we do not have any significant change to report a transition. Consequently, we report counting the noise shows a significant change in the specified region of transition.}
\label{fig:barplot_range comparison}
\end{figure*}

Using the ML model that was trained on the subsampled phase space dataset,  we classify the computed barcodes from the bifurcation parameters (for $\gamma=0.35$ to $\gamma=0.41$). These inputs were not given as input for the train and test set for the classifier. This unseen bifurcation range of parameters encompasses both periodic and chaotic regimes. Based on the computed threshold our trained model classifies the features and noise for the unknown bifurcation parameter range. Our objective is to find whether the classified features and noise could characterize or distinguish the periodic and chaotic regimes. Figure \ref{fig:barplot_features} illustrates the count of features classified by the model for different $\gamma$ values and we mark the transition into chaos by red dashed lines. We observed a fluctuating count of features in the chaotic regime and a relatively stable count in the periodic regime. However, counting the machine-classified features seems trivial as they do not provide a clear distinction during the transition. This inconsistency in counts of features in the chaotic regime arises because of the reduced longevity of features that fall below the machine learning computed threshold, causing them to be classified as noise rather than features. On the contrary, we make a compelling observation by counting the noise in Fig. \ref{fig:barplot_noise}, where we observe a significant increase in the count of noise as we transcend into the chaotic regime. The rationale behind the increasing noise in chaotic regimes is attributed to the non-repeating trajectories in the phase space, a characteristic feature of a chaotic phase. We visually illustrate our saying in Fig. \ref{fig:noise comparison}, this plot compares the barcodes obtained of the periodic regime of $\gamma=0.35$ and chaotic regime $\gamma=0.39$ of the duffing oscillator. We see that the phase portrait of the periodic regime has smooth regular loops that have a repetitive behavior while the phase portrait of the chaotic regime has an irregular and complex structure contributing to noise barcodes. This finding is intriguing because traditionally PH is used to keep track of the long-lived features while disregarding the noise. Conversely, we also got similar results in Fig. \ref{fig:barplot_range comparison} which uses an ML model that was trained on a periodic range of bifurcation parameters for learning. However, we declare that the only difference in opting for this approach (model trained on a periodic range of bifurcation parameters) is the consistency in the model in predicting the count of the true features in the periodic regime. We illustrate this in Fig. \ref{fig:barplot_range_F} where we notice a fluctuating count of features in the chaotic regime while the count of features remains consistent in the periodic regime (unlike the first approach). As expected, the count of noise illustrated in Fig. \ref{fig:barplot_range_N} shows a distinct increase as the system enters the chaotic regime. These findings suggest that counting noise can be a robust quantifier for distinguishing dynamical regimes. Accordingly, there are works in the literature that make use of these topological noises as an active characterization indicator in different applications: the work of Jaquette et al. \cite{jaquette2020fractal} employs PH for dimension estimation in fractals and attractors, notably, they emphasize the importance of considering noise, which can serve as a valuable signal; the work of Maletic et al. \cite{maletic2016persistent} presents a topological framework employing PH in analyzing time series data of dynamical systems using delay-coordinate reconstruction, where they showcase the importance of considering the behavior of topological noise in choosing the optimal embedding dimension parameters. Based on these observations, we propose a methodology that uses topological noise to characterize the dynamical states of systems. The implementation requires a two-pronged approach: using an ML classifier to distinguish between features and noise; and utilizing the noise count to characterize the regimes actively.

Hence, we propose two topological summaries in the form of Persistence Score (PS) and Noise Score (NS) that leverage noise as a crucial parameter for distinguishing system behaviors.  We define PS and NS as:

\begin{equation} \label{PS}
 PS = N/(F + 1) 
\end{equation}
  
\begin{equation} \label{NS}
 NS = N 
\end{equation}

where $N =$ number of noise and $F =$ number of features.  In Eqn.\ref{PS}, the ratio of noise to features is adjusted by adding 1 to the denominator to prevent the ratio from approaching infinity if the model does not detect any feature. This ratio also highlights periodic regimes with higher periodicities with lower scores, thereby enhancing the distinction of chaotic regimes. In Eqn.\ref{NS}, solely relies on the model-classified noise to differentiate the system as justified earlier.

\section{Results:}
We chose three widely studied complex systems to implement our algorithm: The duffing oscillator, a 2-D system, the Lorentz attractor, and the Jerk circuit, a 3-D system. These systems were chosen to showcase the versatility of our algorithm in both 2-D and 3-D systems.

\subsection{Duffing Oscillator:}
\label{duff_section}
\begin{figure}[h]
     \centering
     \begin{subfigure}[b]{0.2\textwidth}
         \centering
         \includegraphics[width=\textwidth]{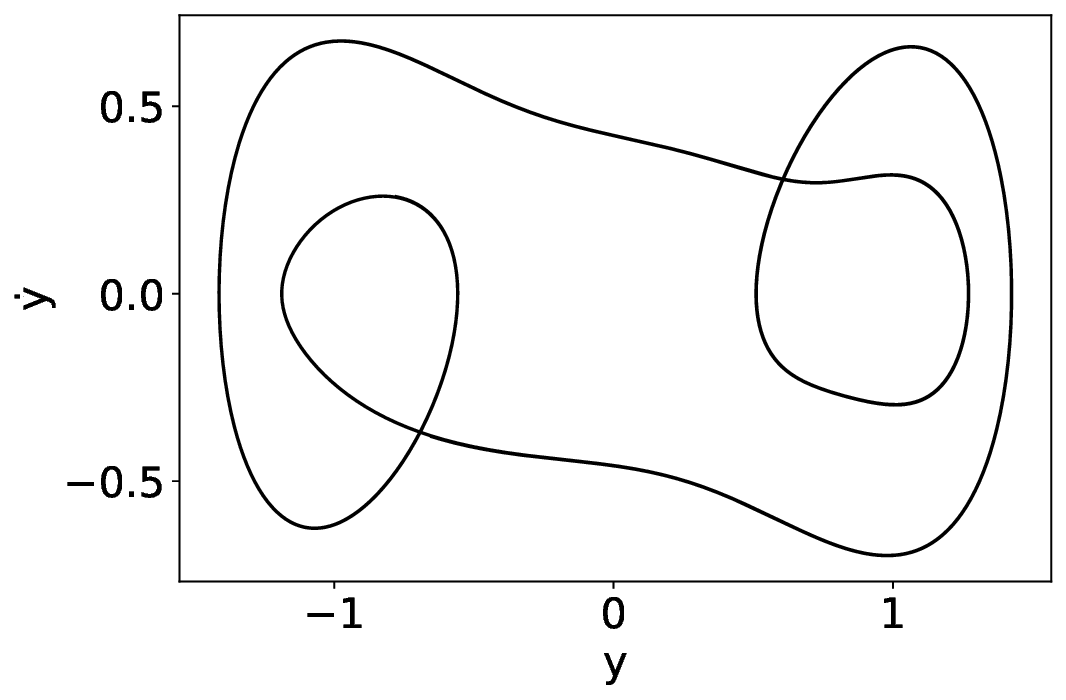}
         \caption{Period-3 oscillation for $\gamma = 0.35$}
         \label{fig:0.35}
     \end{subfigure}
     \begin{subfigure}[b]{0.2\textwidth}
         \centering
         \includegraphics[width=\textwidth]{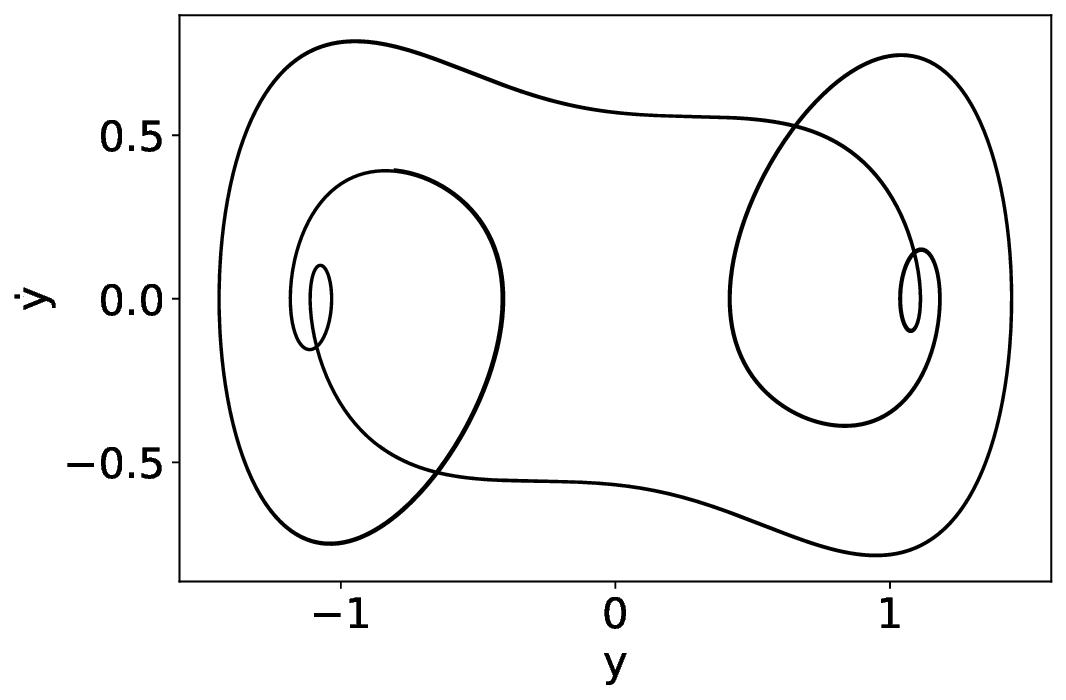}
         \caption{Period-5 oscillation for $\gamma = 0.37$}
         \label{fig:0.37}
     \end{subfigure}
     \begin{subfigure}[b]{0.2\textwidth}
         \centering
         \includegraphics[width=\textwidth]{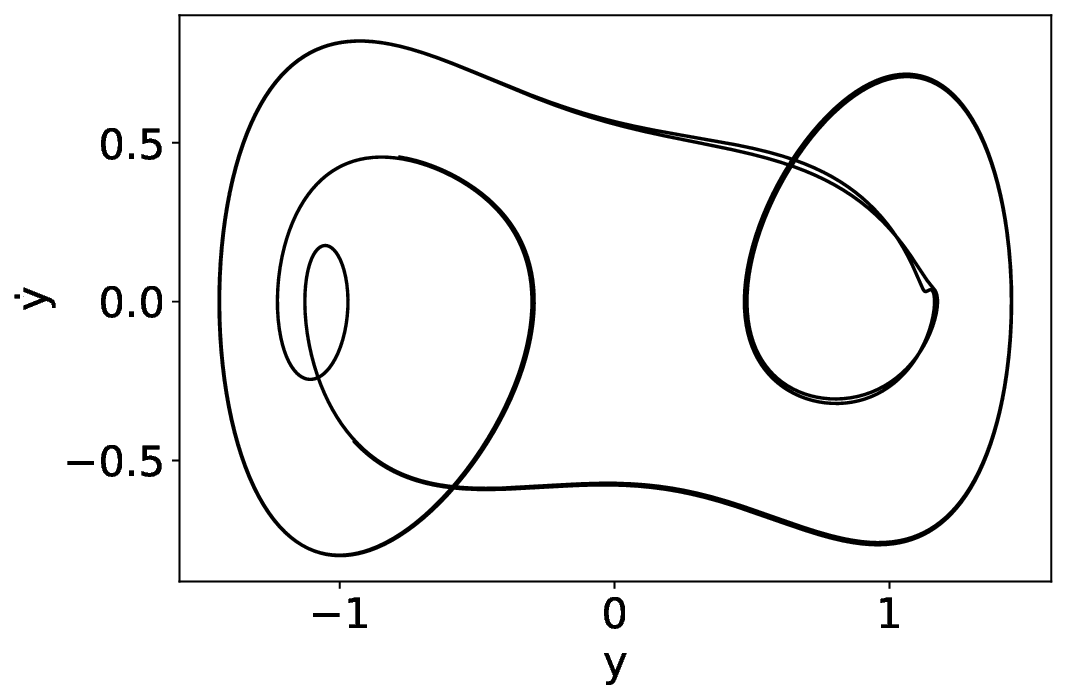}
         \caption{Period-4 oscillation for $\gamma = 0.38$}
         \label{fig:0.38}
     \end{subfigure}  
     \begin{subfigure}[b]{0.2\textwidth}
         \centering
         \includegraphics[width=\textwidth]{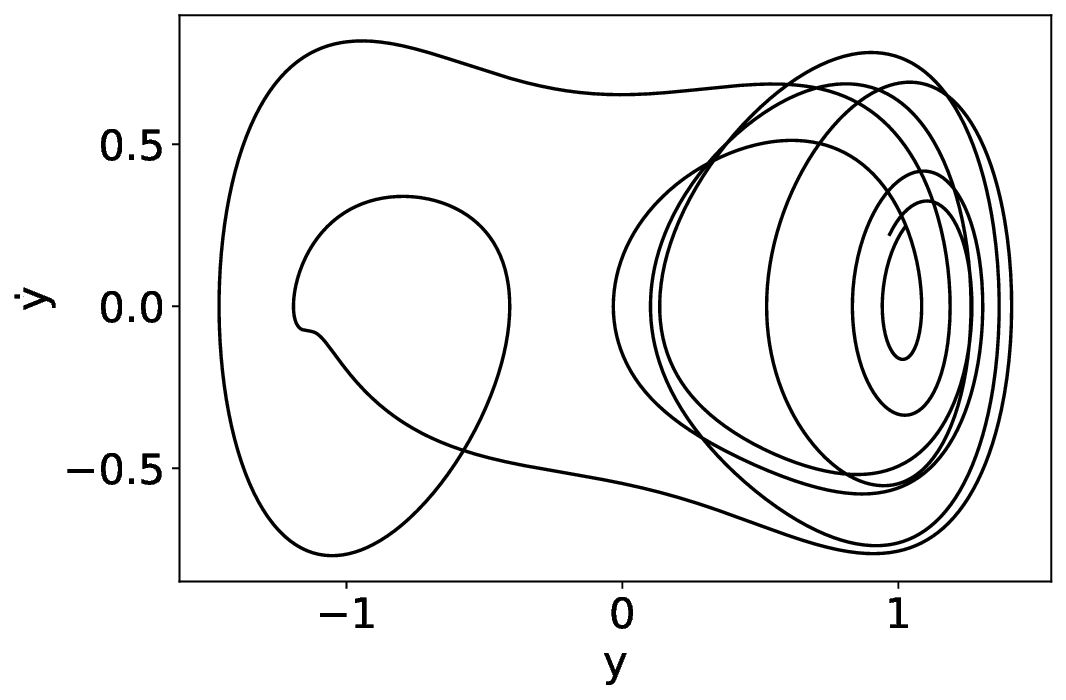}
         \caption{Chaotic phase for $\gamma = 0.40$}
         \label{fig:0.40}
     \end{subfigure}
     	        \caption{Bifurcation states of duffing oscillator obtained by changing the bifurcation parameter $\gamma$. We show within the chosen range of $\gamma$, the system exhibits multiple periodic regimes and eventually transcends into the chaotic regime.}
        		\label{fig:bifurcation}
     \end{figure}

The duffing oscillator is an example of a two-dimensional, non-autonomous, non-linear system that is damped-driven. The duffing oscillator exhibits both 
period-doubling cascade and chaotic behavior due to its non-linear terms.
The equation that governs the motion of the oscillator can be written as: 
$$
\ddot{y}+\delta \dot{y}+\alpha y(t)+\beta y^{3}(t)=\gamma \cos (\omega t)
$$
where, $\gamma$ represent the driving force amplitude and $\delta$ represents the damping coefficient. The system undergoes dynamical changes from periodic to chaotic behavior by changing the parameter $\gamma$ while keeping the other parameters constant at $\beta=1, \alpha=-1, \delta=0.3, \omega=1.2, y(0)=0, \dot{y}(0)=0 $. As shown in the Fig. \ref{fig:bifurcation}, the system showcases period-3 oscillations for $\gamma=0.35$ and $\gamma=0.36$, period-5 oscillations for $\gamma=0.37$ and ends with period-4 oscillations for $\gamma=0.38$. Beyond $\gamma=0.38$ the system exhibits chaotic behavior.

Our primary objective is to predict and classify the dynamical nature of any test phase space data as either periodic or chaotic. The intended method should be to address the challenge wherein only a single data set is available where the data is known to exhibit periodic behavior, and the period of the system is predetermined. This method aims to model situations not uncommon in real-world systems wherein their system parameters are unknown or unmeasurable. Then it becomes imperative to find the dynamical nature of the system as either periodic or chaotic provided their phase space or time series data are made available. In this paper, we confine our study to cases where we have access to the phase space data. We note that it is not difficult to extend this study to cases where we have only the time series data made available instead of the phase space data, in this case, one has to reconstruct the phase space from the time series and employ this procedure. 

Going by our objective, in the Duffing system, we are interested in binary classifying the phase space data as either periodic or chaotic. The machine learning classification method we employ here will essentially have a single input phase space data that is periodic with a known period. This requirement arises because supervised ML relies on labeled data, which implies that the classification of features and noise in barcodes must be manually labeled, which demands prior knowledge of the system's periodicity. We know from the bifurcation diagram of the Duffing system that it exhibits periodic behavior for $\gamma \le 0.38$ and chaotic for $\gamma > 0.38$. Therefore, we can compare the performance of our method with the standard results. 

As we discussed in the section \ref{sec:input generation}, we limit the landmark sizes to necessarily a small landmark size that is randomly sampled. For this analysis, we use 600 landmarks as a representative dataset, although this number can vary depending on different experimental conditions or outcomes. However, our focus remains on addressing the issue of scanty data, rather than the size of the data itself, as this can be subjective. We reiterate (see section \ref{sec:input generation}) that this low number of landmarks is not only an imposed limitation due to the computational handicap but also mimics real-world scenarios where data is scanty. 

\begin{figure}[h]
\centering
\includegraphics[width=0.3\textwidth]{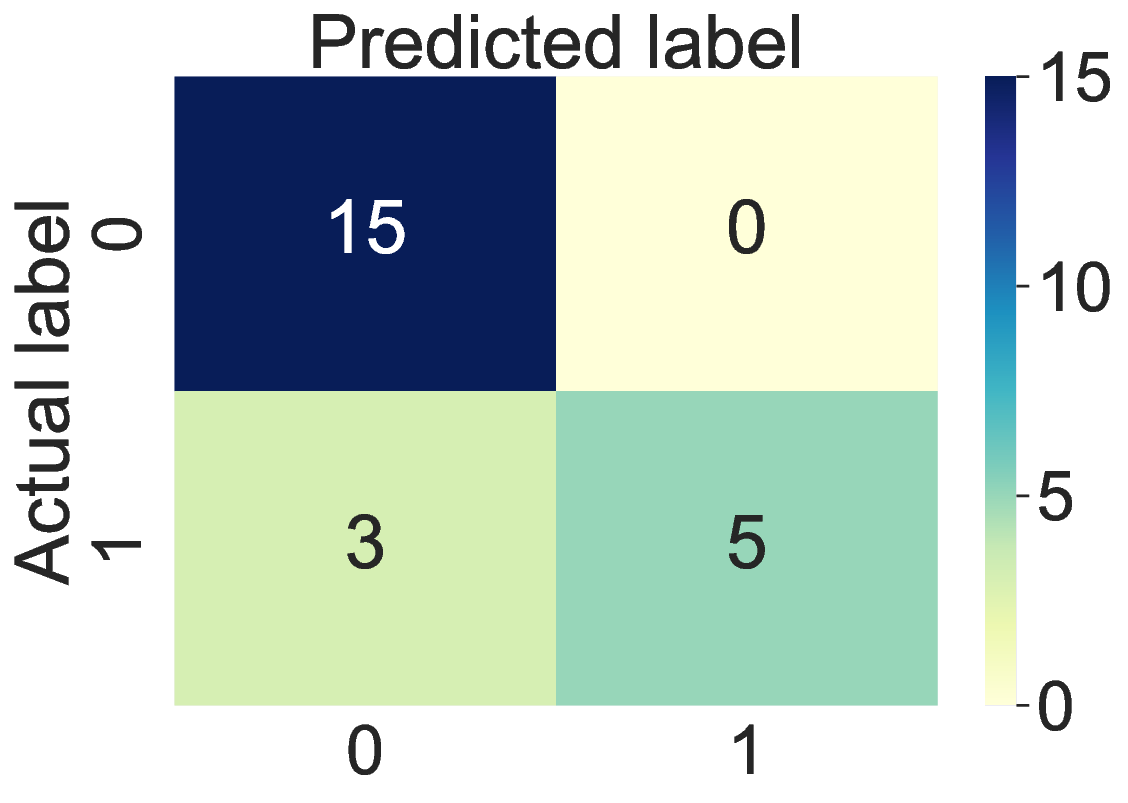}
    \caption{Confusion matrix is evaluated for test data which comes from the $20 \%$ data  (23 test data)  that are randomly taken from the human classified repository of 114 barcode data of the duffing system (for bifurcation parameter $\gamma = 0.35$), while the other $80 \%$ data was used as learning set to train the model. The classified categories obtained from the model represent 15 correctly predicted noise, 0 wrongly predicted noise, and 3 wrongly predicted features, 5 correctly predicted features.}
    \label{fig:cnf_duff}
\end{figure}

We quickly realize that the barcodes obtained by computing PH on the obtained known periodic phase space data of 600 landmarks are too small to train and test an ML classifier effectively. To address this, we use a method (call it \emph{scheme 1}) where we don't require additional input data, but instead we sub-sample the original phase space data to multiple data sets that yield 500, 450, 400, 350, 300, 250, 200, and 100 landmarks; the subsampled data obtained are randomly sampled. Next, we use these multiple data sets and compute their PH individually. Further, manually label the features with the tag `1` and noise with the tag `0`. Finally, we merge all the data sets that are differently sub-sampled into one master data set. This master set is fully tagged and is therefore useful to train and test our ML algorithm and validate its performance. For this purpose, we earmark $80\%$ of the tagged data for training the ML and the remaining $20\%$ of the tagged data for testing its performance. We randomly allocate the training and test tagged data from the master set. The results of our evaluation done for 23 data points ($20\%$ of the tagged data) are presented in Fig. \ref{fig:cnf_duff} wherein we see that the ML has correctly predicted 15 points as noise, 5 points as feature; but wrongly predicted 3 points as feature and 0 points as noise and we achieved an accuracy of $86.9\%$. Now, using the trained classifier, we automatically classify the features and noise for the barcodes obtained by computing PH for other bifurcation parameters: $\gamma = 0.351$ to $\gamma = 0.41$ in steps of $0.001$. We employ the classified features and noise in the proposed topological summaries to characterize the transition.

\begin{figure}
     \centering
     \begin{subfigure}[b]{0.35\textwidth}
         \centering
         \includegraphics[width=\textwidth]{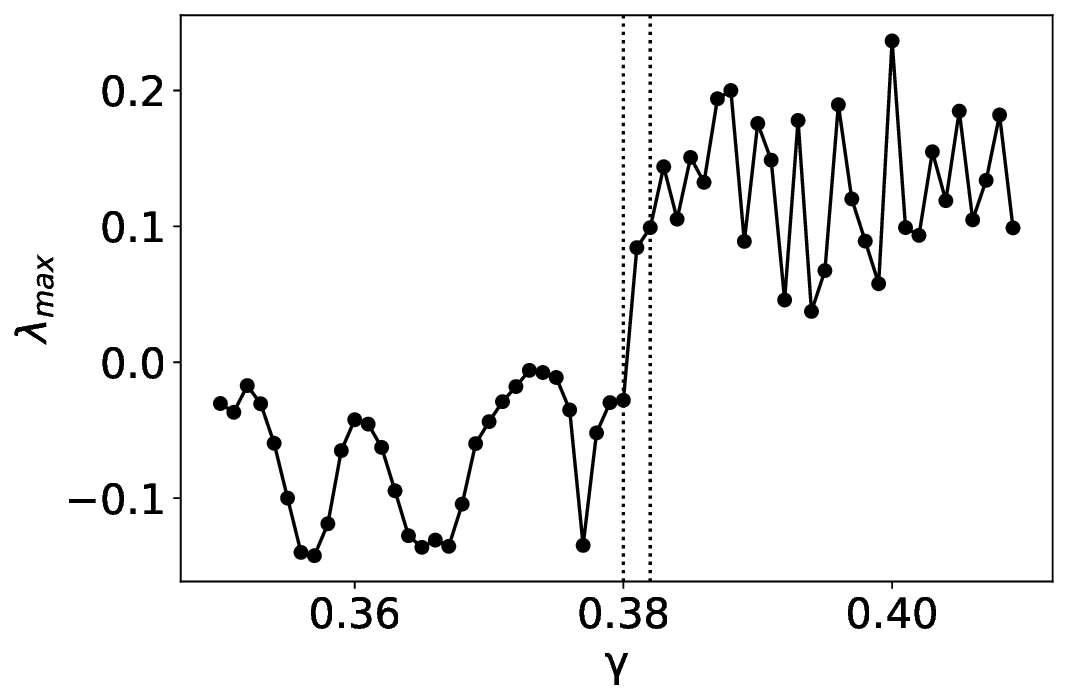}
         \caption{MLE for duffing system}
         \label{fig:MLE}
     \end{subfigure}
     \hfill
     \begin{subfigure}[b]{0.35\textwidth}
         \centering
         \includegraphics[width=\textwidth]{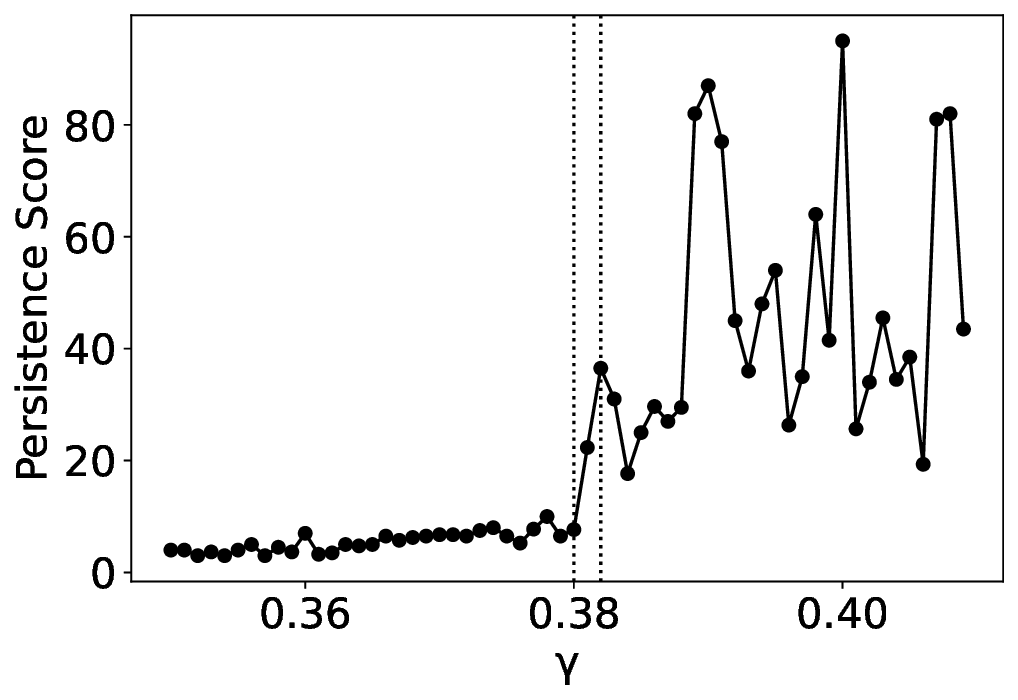}
         \caption{Persistence Score for duffing system}
         \label{fig:PS}
     \end{subfigure}
    \hfill
	 \begin{subfigure}[b]{0.35\textwidth}
         \centering
         \includegraphics[width=\textwidth]{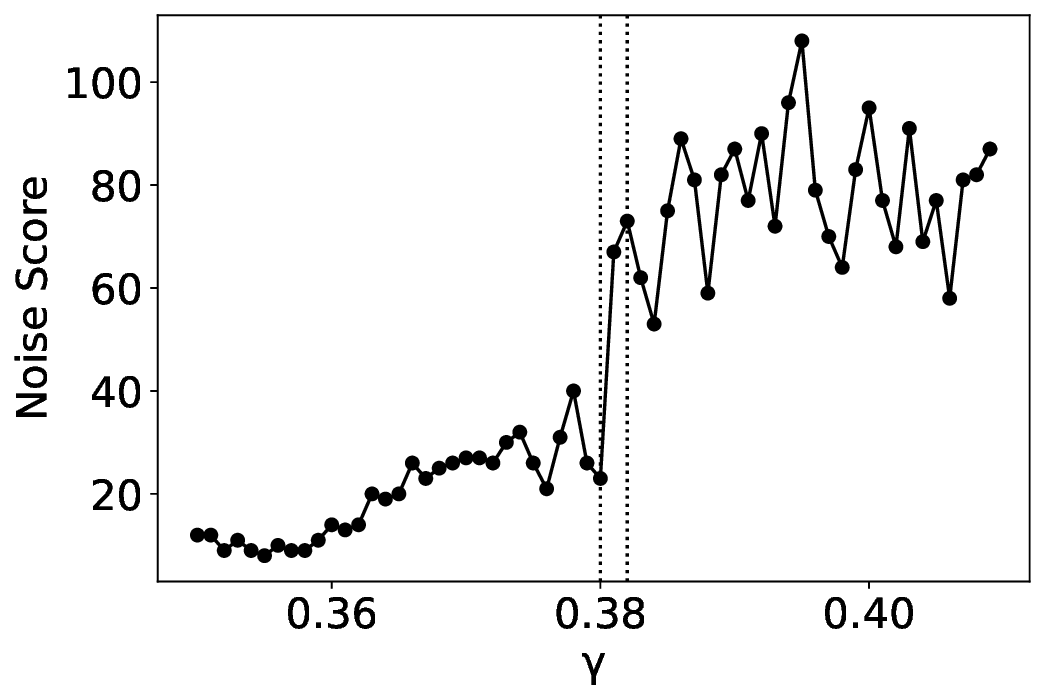}
         \caption{Noise Score for duffing system}
         \label{fig:NS}
     \end{subfigure}
     	        \caption{The plot illustrates the results obtained using the proposed topological summaries: Persistence score and Noise score benchmarked with MLE plot. The transition region is highlighted with dotted points, spanning from $\gamma = 0.38$ to $\gamma = 0.382$ and we showcase that our results match exactly with the MLE.}
        		\label{fig:result_duff}
     \end{figure}

In Fig. \ref{fig:result_duff}, We have our PS (Fig. \ref{fig:PS}) and NS (Fig. \ref{fig:NS}) results benchmarked using the Maximum Lyapunov Exponent (MLE) (Fig. \ref{fig:MLE}).  Notably, the onset of chaotic behavior becomes discernible within the parameter range of $\gamma = 0.38$ to $\gamma = 0.382$ which are marked by dotted vertical lines.

As a control to check the performance of our proposed \emph{scheme 1} wherein the ML algorithm was trained with a single instance of the bifurcation parameter $\gamma$, we employ another method, \emph{scheme 2}, where the ML algorithm is trained with data sets from a range of bifurcation parameters $\gamma$.
As mentioned in section \ref{sec:input generation}, increasing the number of data sets for training ML
(scheme 2) adds a cushioning effect and thereby it provides better classification results, and this should not give surprise to an ML practitioner.
We make use of periodic regime parameters say: $\gamma=0.35$ to $\gamma=0.38$ in steps of 0.1 (without considering the chaotic regimes) as the input for the model. However, this approach requires their respective periods to be known for labeling the barcodes which can be a demanding process.  We compute the PH and classify the features and noise of the barcodes based on their periodicities. We repeat the same procedure that was mentioned above and use the train test split proportion.
Now the trained model is set to classify the barcodes obtained by computing PH for the bifurcation parameters: $\gamma = 0.351$ to $\gamma = 0.41$ in steps of $0.001$. We found that the results were quantitatively similar to the results that were demonstrated in Fig. \ref{fig:result_duff}. This finding supports our approach or the rationale of subsampling the phase space data into multiple smaller sizes of landmarks, reinforcing the idea that it yields similar effects and has negligible impact on the overall results.

\subsection{Lorentz system:}
\begin{figure}[h]
     \centering
     \begin{subfigure}[b]{0.2\textwidth}
         \centering
         \includegraphics[width=\textwidth]{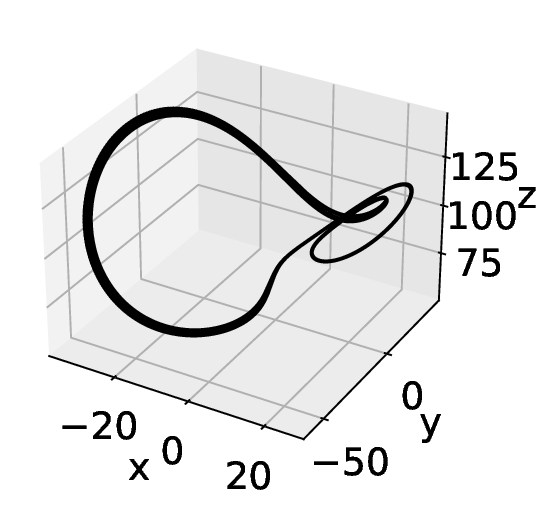}
         \caption{Periodic oscillation for $\rho = 99.56$}
         \label{fig:160}
     \end{subfigure}
     \begin{subfigure}[b]{0.2\textwidth}
         \centering
         \includegraphics[width=\textwidth]{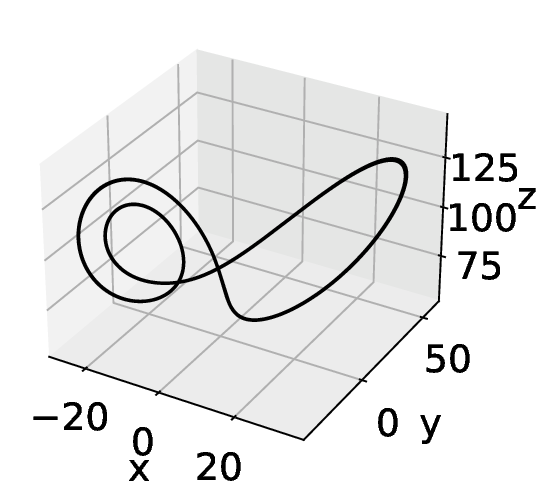}
         \caption{Periodic oscillation for $\rho = 100$}
         \label{fig:132}
     \end{subfigure}  
     \begin{subfigure}[b]{0.2\textwidth}
         \centering
         \includegraphics[width=\textwidth]{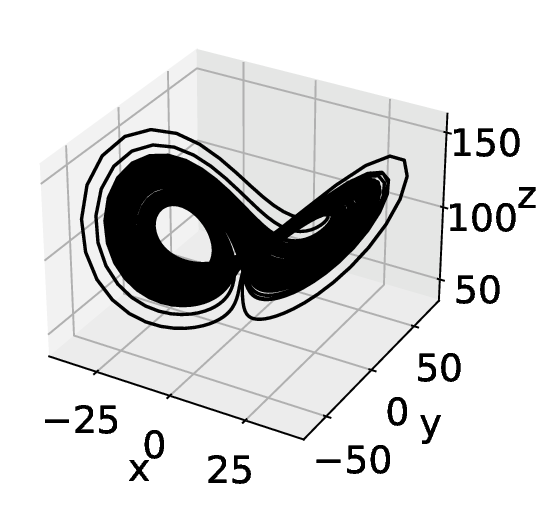}
         \caption{Chaotic phase for $\rho = 101$}
         \label{fig:100}
     \end{subfigure}
	 \begin{subfigure}[b]{0.2\textwidth}
         \centering
         \includegraphics[width=\textwidth]{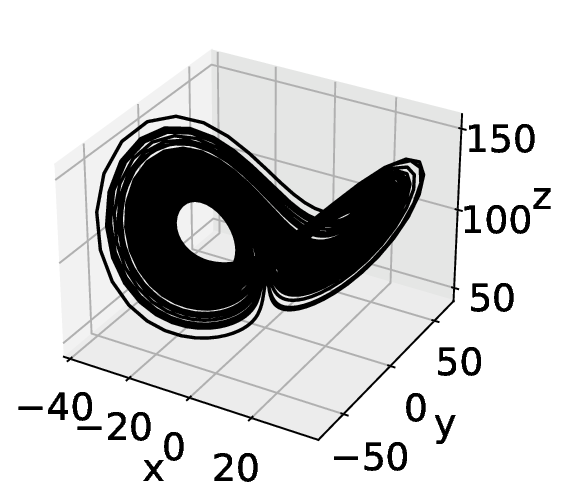}
         \caption{Chaotic phase for $\rho = 102$}
         \label{fig:95}
     \end{subfigure}
     	        \caption{Bifurcation states of Lorentz oscillator obtained by changing the bifurcation parameter $\rho$. We show within the chosen range of $\rho=99.56$ to $\rho=102$, we have distinctive periodic regimes and transcends into the chaotic regime.}
        		\label{fig:bifurcation_lorentz}
     \end{figure}

In 1963, Edward Lorentz simplified a 12-dimensional system for modeling atmospheric convection to a set of 3 ordinary differential equations known as Lorentz equations. It is a three-dimensional, nonlinear, aperiodic, deterministic system that exhibits both periodic and chaotic phases for different bifurcation parameters. The Lorentz equations are given by:
$$\dot{x} = \sigma (y - x)$$\\
$$\dot{y} = x(\rho - z) - y$$
$$\dot{z} = xy - \beta z$$
Bifurcation occurs by changing the bifurcation parameter rho ($\rho$) and keeping the other parameters constant $\sigma =10$, $\beta = 8/3$, $x(0) = 1$, $y(0) = 1$, $z(0) = 1$. As shown in Fig. \ref{fig:bifurcation_lorentz}, the system showcases period-3 for $\rho=99.56$ as it moves towards $\rho=100.8$ it shows the transition into the chaotic phase. 

\begin{figure}[h!]
\includegraphics[width=0.3\textwidth]{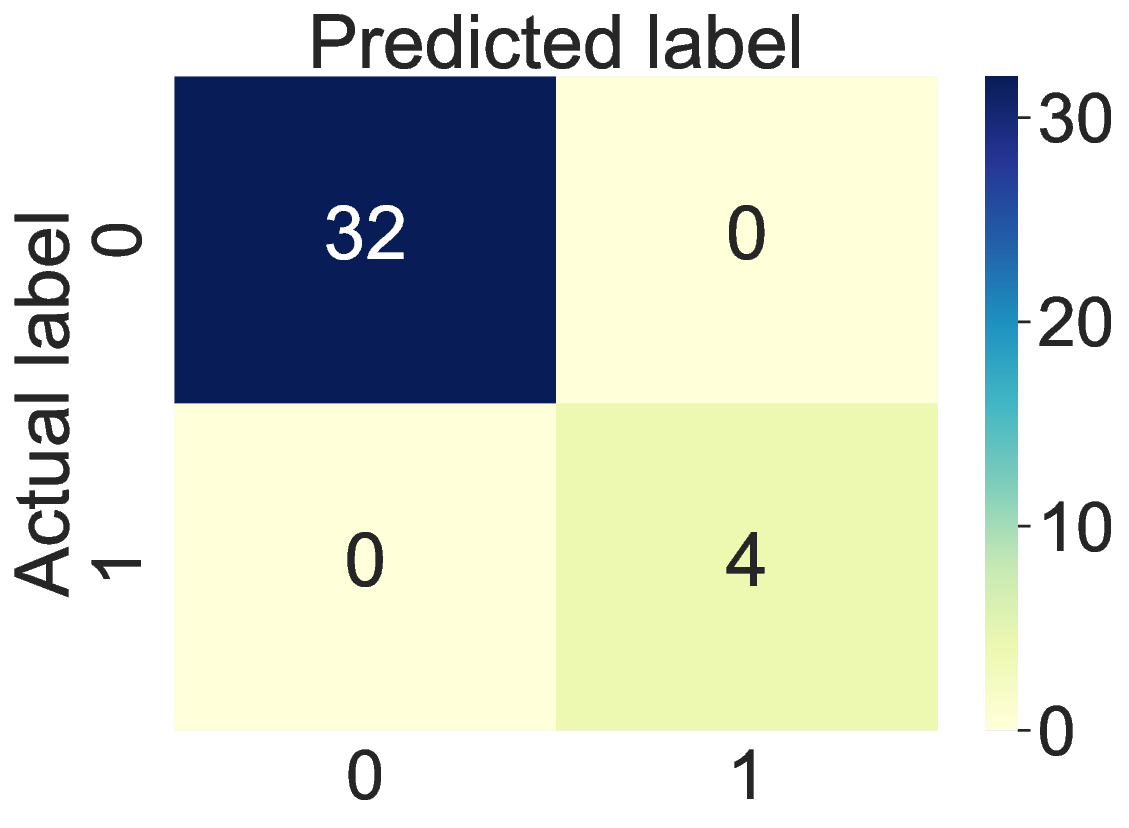}
    \caption{Confusion matrix was evaluated using test data, which comprised $20\%$ of the dataset (36 samples) randomly selected from a manually classified repository of 179 barcodes dataset. The classification results indicated that the model correctly predicted 32 noise, 0 wrongly predicted noise, and 0 wrongly predicted features, 4 correctly predicted features.}
    \label{fig:cnf_lor}
\end{figure}

We begin with a scanty data of 600 landmarks from a periodic regime for $\rho = 99.56$ which is period-3, as our given input with apriori information.
We compute PH to extract the associated barcodes and employ the method (\emph{scheme 1} we had used for the Duffing system (refer \ref{duff_section}) to generate
sub-sampled data sets from the original landmark data. 
That is, we sub-sampled the given 600 landmark data into 500, 400, 300, 200, and 100 landmark sizes and computed the PH on them separately to extract their barcodes. We manually label the features as tag `1' and the noise as tag `0' based on their period of oscillation (we have considered a period 3 system, hence there should be 3 features). This process resulted in a total of 179 barcodes, allowing us to gather sufficient data for further analysis. We use a $80\%$ training set and $20\%$ test set split for the classifier. We present the evaluation results for the test set performed on 36 barcodes using the confusion matrix as illustrated in Fig. \ref{fig:cnf_lor} where we have 32 correctly predicted noise (0), 4 correctly predicted feature (1), and 0 wrongly predicted features and noise. Hence, this yields an accuracy of $100 \%$. Now, using the trained classifier model we automate the process of classifying the features and noise for other bifurcation parameters from $\rho = 99.56$ to $\rho = 102$ in steps of $0.02$. We use the classified features and noise in the topological summaries to characterize the transition.

\begin{figure}[h!]
     \centering
     \begin{subfigure}[b]{0.35\textwidth}
         \centering
         \includegraphics[width=\textwidth]{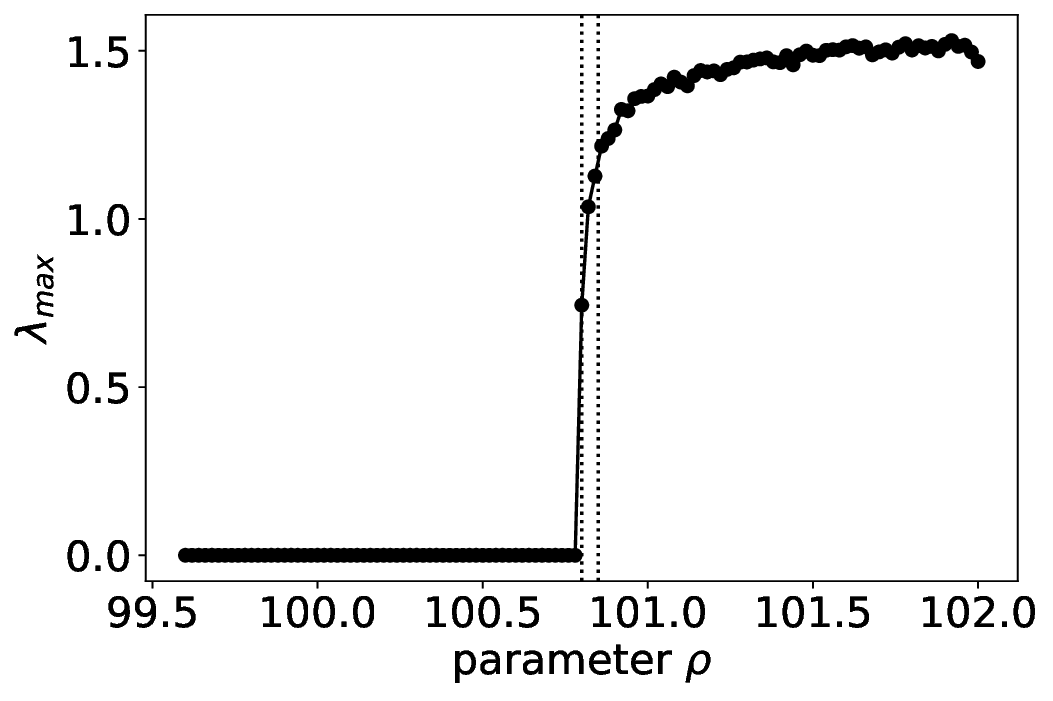}
         \caption{MLE for Lorentz system}
         \label{fig:lor_MLE}
     \end{subfigure}
     \hfill
     \begin{subfigure}[b]{0.35\textwidth}
         \centering
         \includegraphics[width=\textwidth]{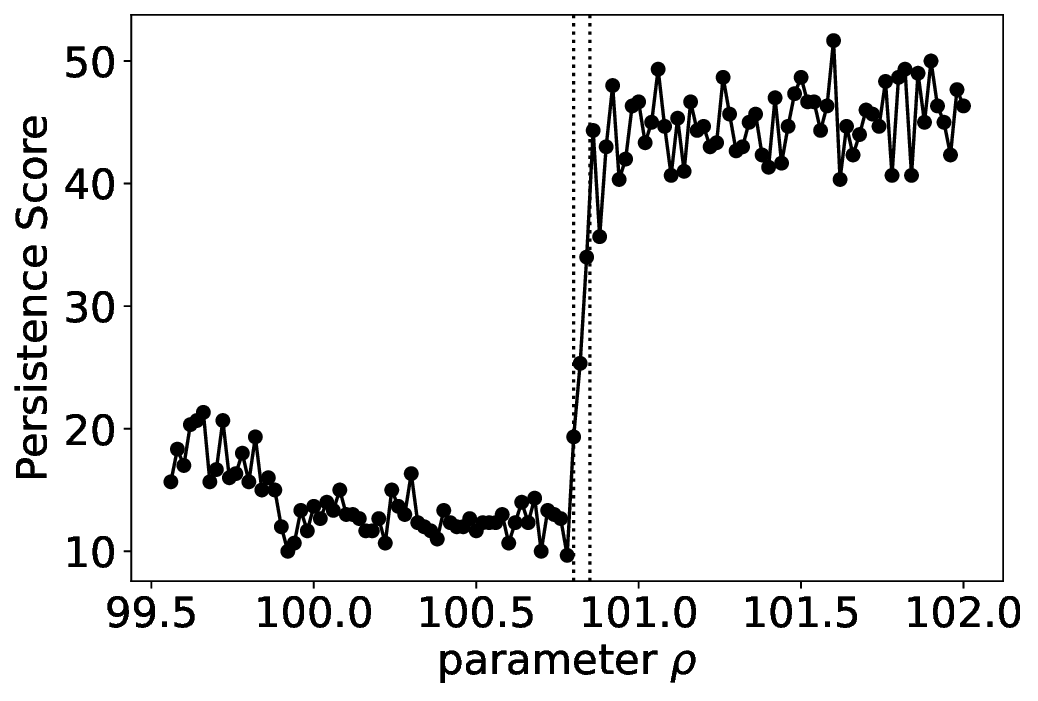}
         \caption{Persistence Score for Lorentz system}
         \label{fig:lor_PS}
     \end{subfigure}
    \hfill
	 \begin{subfigure}[b]{0.35\textwidth}
         \centering
         \includegraphics[width=\textwidth]{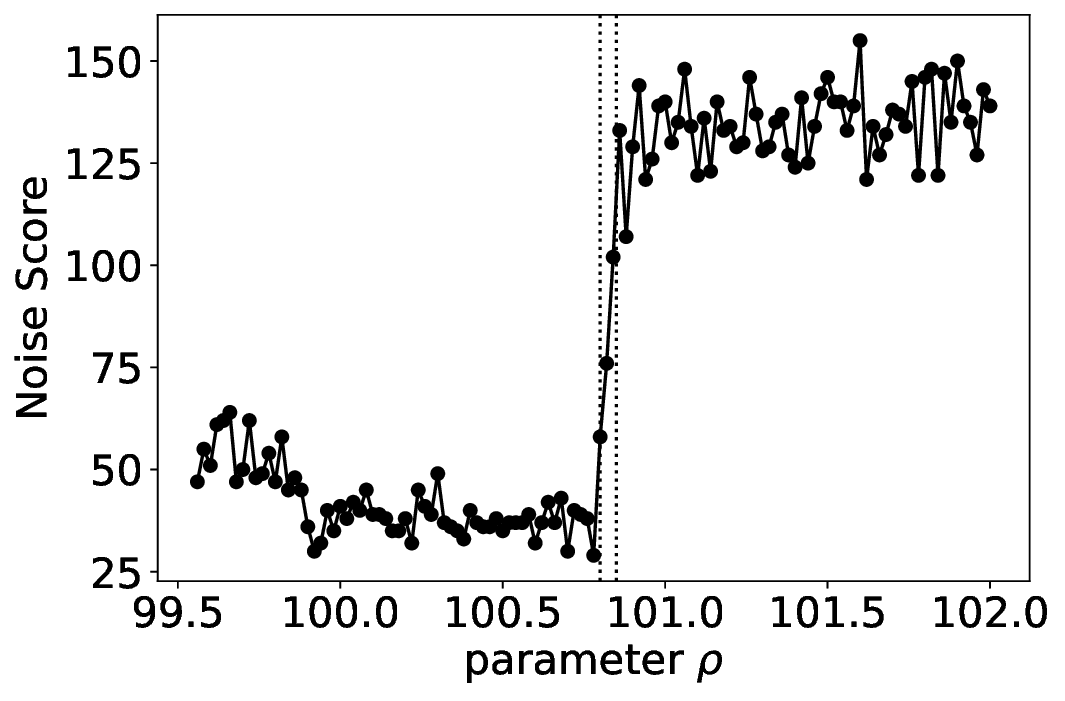}
         \caption{Noise Score for Lorentz system}  
         \label{fig:lor_NS}
     \end{subfigure}
     	        \caption{The plot illustrates the results obtained using the proposed topological summaries: Persistence score and Noise score benchmarked with the MLE plot. The transition region is highlighted with dotted points, spanning from $\rho = 100.8$ to $\rho = 100.85$.  In this instance, our results for PS and NS exhibit similar quantitative behavior.}
        		\label{fig:result_lor}
     \end{figure}

In Fig. \ref{fig:result_lor}, We have our PS (Fig. \ref{fig:lor_PS}) and NS (Fig. \ref{fig:lor_NS}) plot matched with the Maximum Lyapunov Exponent (MLE) (Fig. \ref{fig:lor_MLE}).  Notably, the onset of chaotic behavior becomes significant within the parameter range of $\rho = 100.8$ to $\rho = 100.85$.

As before, we also employ the control measurements with \emph{scheme 2} by using periodic regime parameters from $\rho = 99.56$ to $\rho = 100.8$ in steps of 0.01 as input data for the classifier, assuming that the a priori period of oscillation for these parameters is known. We repeat the same procedure to train and test obtained by computing PH for the scanty phase space data of 600 landmarks for the above-mentioned range of parameters. Now using the trained model we try to classify the computed barcodes obtained from bifurcation parameters from $\rho = 99.56$ to $\rho = 102$ in steps of $0.02$. The topological summaries obtained were quantitatively similar to the ones obtained in Fig. \ref{fig:result_lor}.

\subsection{Jerk circuit:}
\begin{figure}[h!]
     \centering
     \begin{subfigure}[b]{0.2\textwidth}
         \centering
         \includegraphics[width=\textwidth]{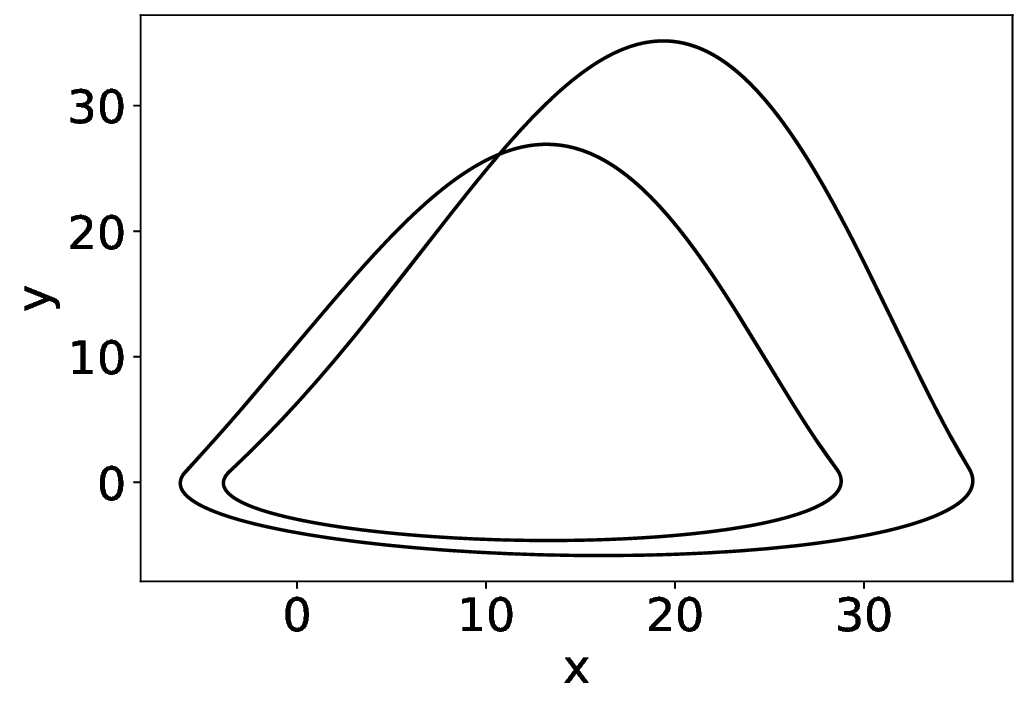}
         \caption{Period-2 oscillation for $\gamma = 0.154$}
         \label{fig:0.154}
     \end{subfigure}
     \begin{subfigure}[b]{0.2\textwidth}
         \centering
         \includegraphics[width=\textwidth]{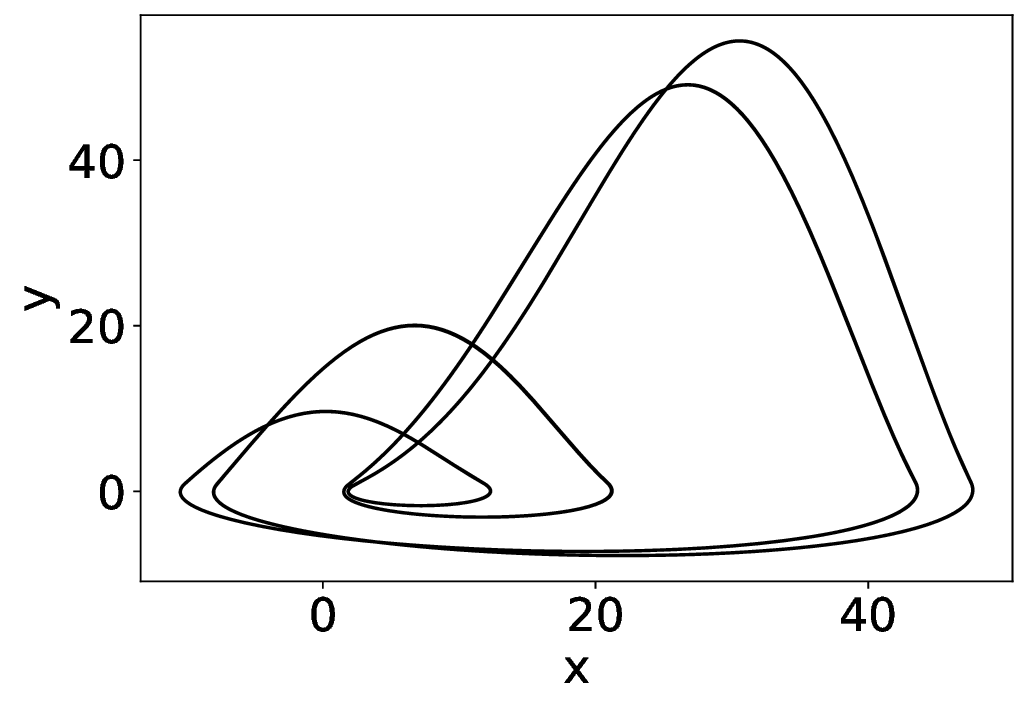}
         \caption{Period-4 oscillation for $\gamma = 0.135$}
         \label{fig:0.135}
     \end{subfigure}
     \begin{subfigure}[b]{0.2\textwidth}
         \centering
         \includegraphics[width=\textwidth]{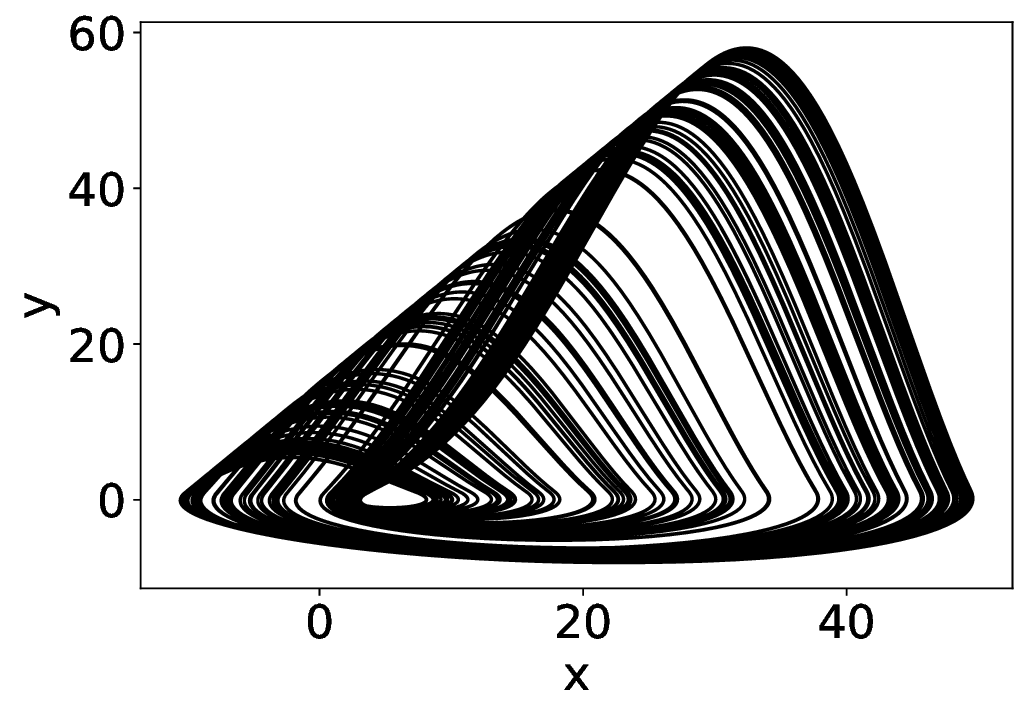}
         \caption{Chaotic phase for $\gamma = 0.117$}
         \label{fig:0.117}
     \end{subfigure}
	 \begin{subfigure}[b]{0.2\textwidth}
         \centering
         \includegraphics[width=\textwidth]{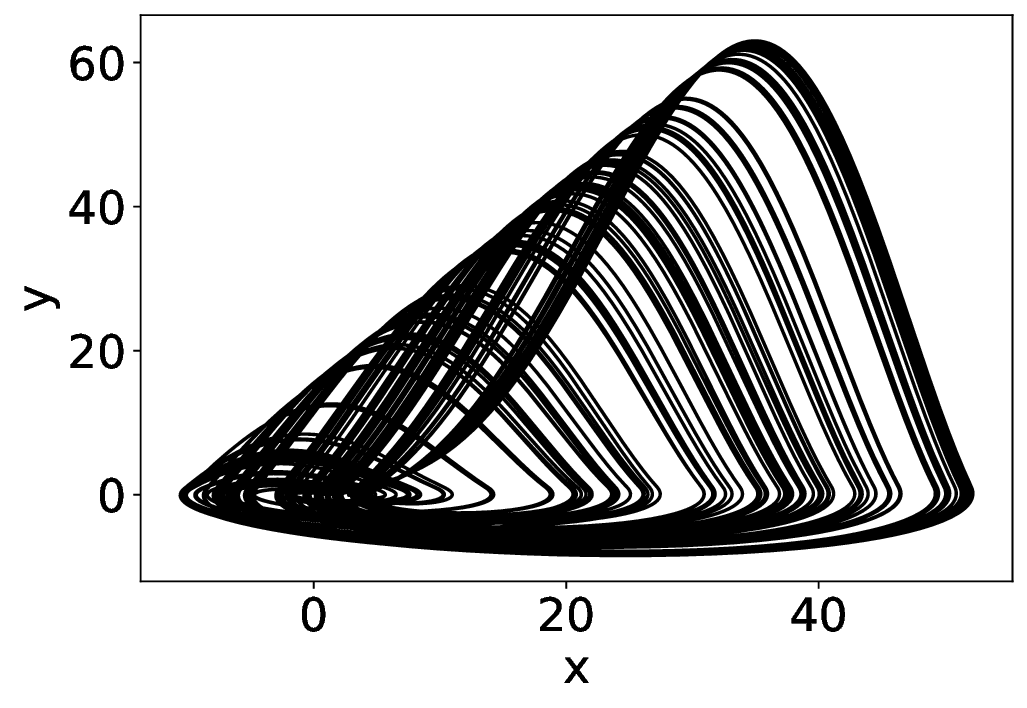}
         \caption{Chaotic phase for $\gamma = 0.11$}
         \label{fig:0.11}
     \end{subfigure}
     	        \caption{Bifurcation states of Jerk system obtained by changing the bifurcation parameter $\gamma$. We show within the chosen range of $\gamma=0.160$ to $\gamma=0.110$, we have multiple periodicities in the periodic regime and transcends into the chaotic regime.}
        		\label{fig:bifurcation_jerk}
     \end{figure}

Jerk circuit is a three-dimensional system represented as $\dddot{x} = J(\ddot{x},\dot{x},x)$, the term Jerk comes from the fact that it is a third-order derivative of displacement ($x$). Jerk circuit exhibits both period-doubling cascade and chaotic behavior which helps us to visualize the onset of chaos. Jerk equation is defined using a state vector $F = [x,y,z]^T$ as:

$$
\frac{dF}{dt} =
	\begin{bmatrix} 
	0 & 0 & 0 \\        
	0 & 0 & -\alpha\\
	\beta & -\beta & -\beta \gamma\\
	\end{bmatrix}
	\quad F \:+	\begin{bmatrix} 
	1 \\
	\alpha \\
	0 \\
	\end{bmatrix}
	\quad G(y)
$$

here, $\beta$, $\gamma$, $\alpha$ are system parameters and G(y) is piece wise non-linear function which is given by:
$$
G(y) =
    \begin{cases}
        -y & \text{if } y \le 1 \\
        -1 & \text{if } y > 1
    \end{cases}
$$
Bifurcation occurs by changing the bifurcation parameter $\gamma$ and keeping the other parameters constant $\alpha =0.025$, $\beta = 0.765$, $x(0) = 0.1$, $y(0) = -5 \times 10^{-7}$, $z(0) = -1$. As shown in Fig. \ref{fig:bifurcation_jerk}, the system showcases period-2 for $\gamma=0.154$ and period-3 for $\gamma=0.13$ and then the chaotic phase from $\gamma=0.121$.

We begin with a scanty data of 600 landmarks from a periodic regime for $\gamma = 0.154$ which is period 2 as our given input with apriori information of its period. We extract the barcodes of the data using PH and using the \emph{scheme 1} we sub-sample it into 600, 500, 400, 300, and, 200 landmark sizes. We manually label the feature and noise barcodes as 1 and 0 respectively based on the period of oscillation. In this case, we label two long persisting features as tag `1' and label others as tag `0'. We compiled the labeled barcodes together which yielded a total of 180 barcodes, sufficiently large data for the ML to work with. We use a $80\%$ training set and $20\%$ test set split for the classifier model. We present the evaluation metrics for the test set performed for 36 barcodes ($20\%$ test set of 180 barcode data) using the confusion matrix as illustrated in Fig. \ref{fig:cnf_lor} where we have 34 correctly predicted noise (0), 2 correctly predicted feature (1), and 0 wrongly predicted feature and noise. We achieved an accuracy of $100 \%$. Now, using the trained classifier model we automate the process of classifying the barcodes into features and noise for other bifurcation parameters from $\gamma = 0.160$ to $\gamma = 0.110$ in steps of $0.001$. We employ the classified features and noise in the topological summaries to characterize the transition.
\begin{figure}
\includegraphics[width=0.3\textwidth]{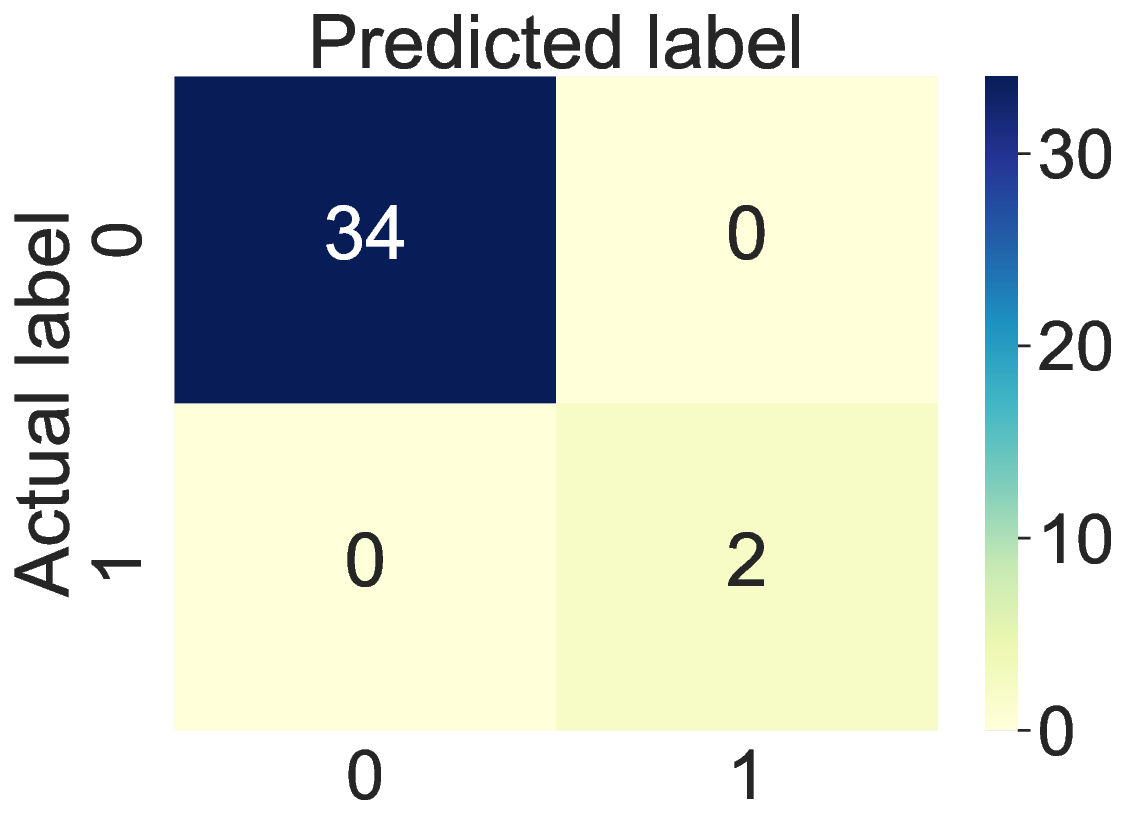}
    \caption{Confusion matrix was evaluated using test data, which comprised $20\%$ of the dataset (36 samples) randomly selected from a manually classified repository of 179 barcodes dataset. The matrix represents 34 correctly predicted noise, 0 wrongly predicted noise, and 2 correctly predicted features, 0 wrongly predicted features.}
    \label{fig:cnf_jerk}
\end{figure}
\begin{figure}[h]
     \centering
     \begin{subfigure}[b]{0.37\textwidth}
         \centering
         \includegraphics[width=\textwidth]{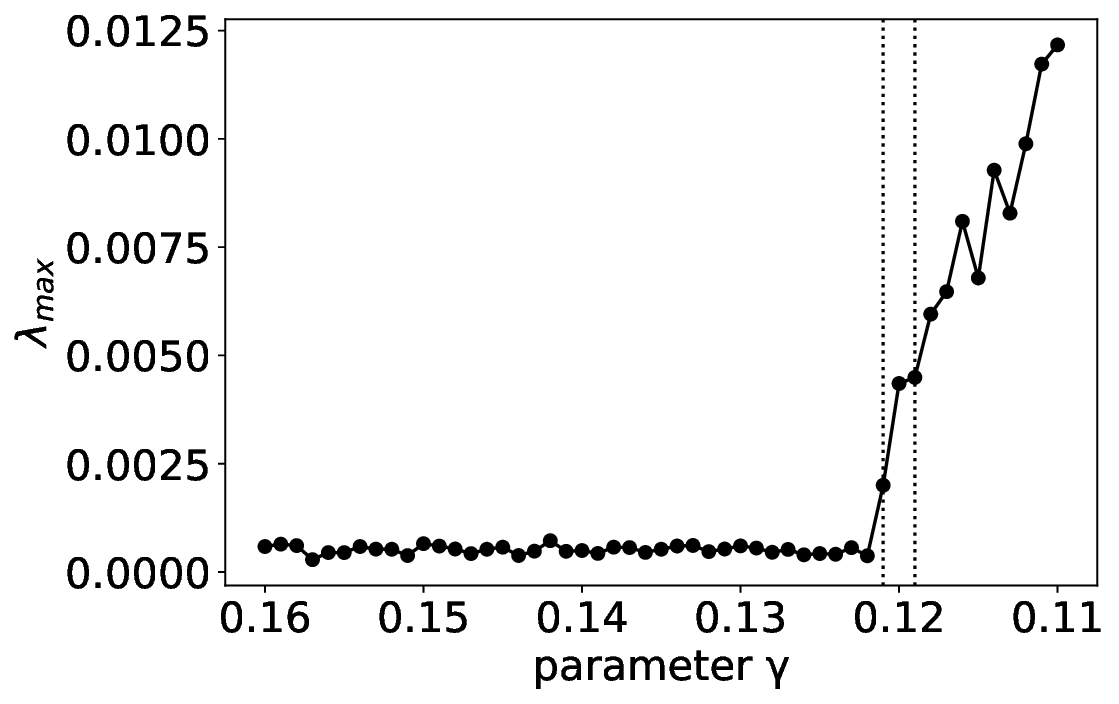}
         \caption{MLE for Jerk system}

         \label{fig:jerk_MLE}
     \end{subfigure}
     \hfill
     \begin{subfigure}[b]{0.35\textwidth}
         \centering
         \includegraphics[width=\textwidth]{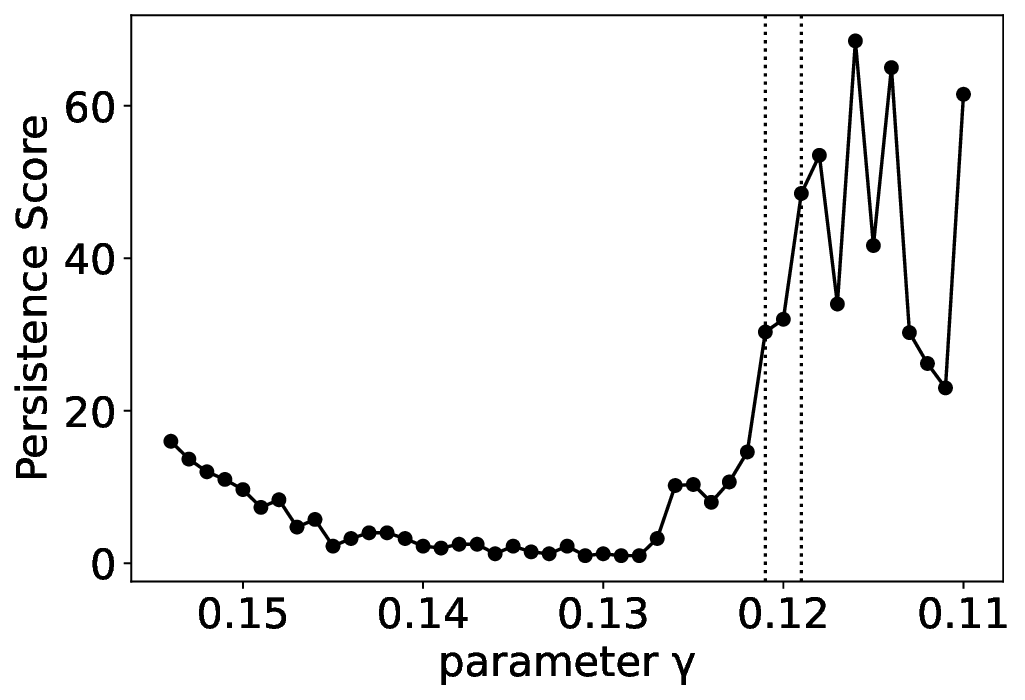}
         \caption{Persistence Score for Jerk system}
         \label{fig:jerk_PS}
     \end{subfigure}
    \hfill
	 \begin{subfigure}[b]{0.35\textwidth}
         \centering
         \includegraphics[width=\textwidth]{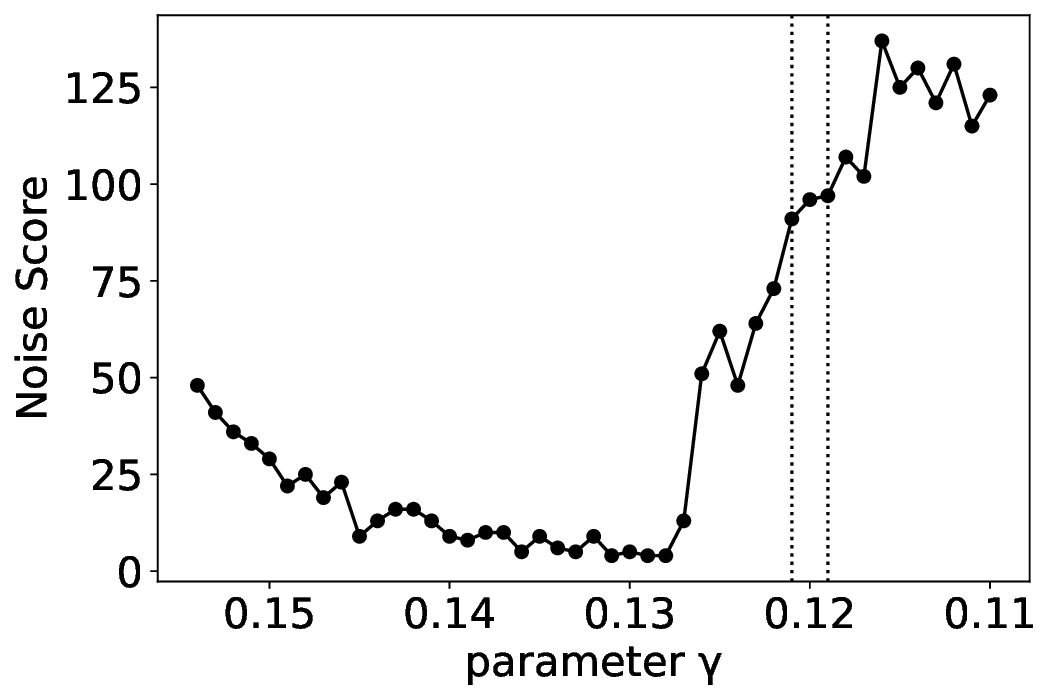}
         \caption{Noise Score for Jerk system}
         \label{fig:jerk_NS}
     \end{subfigure}
     	        \caption{The plot illustrates the results obtained using the proposed topological summaries: Persistence score and Noise score benchmarked with the MLE plot. The transition region is highlighted with dotted points, spanning from $\gamma=0.119$ to $\gamma = 0.121$.}
        		\label{fig:result_jerk}
     \end{figure}

In Fig. \ref{fig:result_jerk}, We have our PS (Fig. \ref{fig:jerk_PS}) and NS (Fig. \ref{fig:jerk_NS}) plot matched with the Maximum Lyapunov Exponent (MLE) (Fig. \ref{fig:jerk_MLE}).  Notably, the onset of chaotic behavior becomes significant within the parameter range of $\gamma = 0.119$ to $\gamma = 0.121$.

Subsequently, as done earlier,  we do a control measurement with \emph{scheme 2} using periodic regime parameters of $\gamma = 0.121$ to $\gamma = 0.160$ in steps of 0.001 as input data for the classifier. This approach demands the period of oscillation for these parameters to be known. We repeat the same procedure to train and test the barcodes obtained by computing PH for the scanty phase space data of 600 landmarks for the above-mentioned range of parameters. Now using the trained model we try to classify the computed barcodes for bifurcation parameters ranging from $\gamma = 0.160$ to $\gamma = 0.110$ in steps of $0.001$ to find the transtioning region. The topological summaries obtained were quantitatively similar to the ones obtained in Fig. \ref{fig:result_jerk}.


\section{Conclusion and Discussion:}

In this paper we present a methodology to classify the periodic and chaotic phase in dynamical systems from scanty or limited phase space data, leveraging tools from topological data analysis and machine learning. Since changing subcycles are active indicators in detecting the bifurcation route to chaos, we use persistent homology to analyze the shape of the data and extract topological features via barcodes. These topological features are identified by focusing on the barcodes that persist over time (long-lived barcodes). However, we observed that validating the true features (long-lived barcodes) becomes challenging without looking at the phase space data and validating it manually. To address this, we use a machine learning-based binary classification algorithm to distinguish true features from noise. We train the ML-based binary classifier model using two schemes. In the first method, \emph{scheme 1}, the ML uses a single input with landmarks 
that are phase space coordinates from a periodic regime of the system corresponding to a single value of $\gamma$. Scheme 1 generates multiple data sets that consist of different landmark sizes by subsampling and computes its barcodes. We manually label the barcodes as features and noise to compile those into a single input file. The second method, \emph{scheme 2}, serves as a control method where the ML uses input that does not originate from a single phase space data and instead considers phase space coordinates from a range of bifurcation parameters in the periodic regime. We compute PH and obtain the barcodes separately by manually labeling the barcodes as features and noise to compile them into a single input file. However, the former method, \emph{scheme 1} just requires apriori information of the period of oscillation for input one phase space data while the latter method, \emph{scheme 2} requires apriori information of the period of oscillation for the range of bifurcation parameters under consideration to label them. Our primary objective is to use \emph{scheme 1} considering that it is simple and needs minimal apriori data for the ML model. We further, train and test the model using the input file generated, and evaluate the model using the confusion matrix and accuracy scores.
Now, the trained model is used to classify the true features and noise for the barcodes obtained from unseen phase space data that may arise by varying the value of bifurcation parameters. 

This exercise also serves as a good model framework to test other real-world systems wherein one has access to only a scanty data set and is tasked to classify the dynamical nature of the system with the variation of system parameters. We plot the counts of classified features and noise obtained across different bifurcation parameters to explore changes in the transition region leading to a chaotic regime. Upon investigation, we observe a noteworthy pattern: a significant increase in the noise count as the system transitions from a periodic to a chaotic state. This telltale role of the noise count in detecting system transitions is surprising and has not been reported before. This is important
because the practitioners of persistent homology look for long-lived features to infer the state of a system and discard the short-lived features. 
But we observe that the short-lived features too can be employed as a differentiating factor in characterizing the regimes of the dynamical systems. This prompts us to make use of the noise count as an active indicator for our analysis. In summary, we devised two topological summaries: persistence score (PS) and noise score (NS) to classify the periodic and chaotic regimes.

The methodology was tested on 2-D systems like the Duffing system and 3-D systems like the Lorenz and Jerk systems. Additionally, we benchmark our proposed topological summaries against the Maximum Lyapunov Exponent (MLE). The results obtained show a clear distinction between the chaotic regime from the periodic regime, aligning well with phase transitions indicated by MLE. All the analyses were done on scanty phase space data to mimic the real-world or experimental data. However, we disclose that in scenarios where measuring any of the state variables is possible or made available, one can opt to reconstruct the time series to get the phase space and proceed with our methodology. We tabulate the evaluation metrics for all three systems in Table \ref{tab:table_acc} to showcase the performance of the model. We hope that this ML-assisted methodology will help in anomaly detection, preventing the systems from further deterioration. Also, we hope that this methodology will be of help to characterize experimental data that are limited and scanty. Further, we expect the current study will be of interest to tackle real-world problems that undergo phase transitions, for instance, the characterization of Atrial fibrillation \cite{jiang2022topological}, chatter in mechanical gadgets \cite{khasawneh2016chatter}, and dynamic state detection in complex networks\cite{myers2019persistent}. 

In future, we plan to extend this work to investigate the role of noise in the systems, which will be valuable for practitioners working on noise-laden experimental systems.

\begin{table}[h!]
\begin{tabular}{|l|l|}
\hline
                        & \textbf{Accuracy (\%)} \\ \hline
\textbf{Duffing system} & 86.9                   \\ \hline
\textbf{Lorentz system} & 100                    \\ \hline
\textbf{Jerk system}    & 100                    \\ \hline
\end{tabular}
\caption{Table representing the accuracy for the Duffing, Lorentz, and Jerk systems respectively. The obtained high accuracy implies the ability of the model to distinguish the features and noise.}
\label{tab:table_acc}
\end{table}
\section{Acknowledgements:}
We wish to thank Mr. Joel Prakash Stephen who initiated this work and laid the foundation as a project student. We extend our gratitude to Dr. Balachandra Suri for early discussions and for introducing us to the reference  \cite{jaquette2020fractal}. 
\section{Conflict of interest:}
On behalf of all authors, we state that there is no conflict of interest

\nocite{*}

\bibliography{myref}

\begin{thebibliography}{10}

\bibitem{khasawneh2016chatter}
Firas~A Khasawneh and Elizabeth Munch.
\newblock Chatter detection in turning using persistent homology.
\newblock {\em Mechanical Systems and Signal Processing}, 70:527--541, 2016.

\bibitem{sujith2020complex}
Raman~I Sujith and Vishnu~R Unni.
\newblock Complex system approach to investigate and mitigate thermoacoustic instability in turbulent combustors.
\newblock {\em Physics of Fluids}, 32(6), 2020.

\bibitem{garfinkel1997quasiperiodicity}
Alan Garfinkel, Peng-Sheng Chen, Donald~O Walter, Hrayr~S Karagueuzian, Boris Kogan, Steven~J Evans, Mikhail Karpoukhin, Chun Hwang, Takumi Uchida, Masamichi Gotoh, et~al.
\newblock Quasiperiodicity and chaos in cardiac fibrillation.
\newblock {\em The Journal of clinical investigation}, 99(2):305--314, 1997.

\bibitem{edelsbrunner2008persistent}
Herbert Edelsbrunner, John Harer, et~al.
\newblock Persistent homology-a survey.
\newblock {\em Contemporary mathematics}, 453(26):257--282, 2008.

\bibitem{edelsbrunner2002topological}
Edelsbrunner, Letscher, and Zomorodian.
\newblock Topological persistence and simplification.
\newblock {\em Discrete \& Computational Geometry}, 28:511--533, 2002.

\bibitem{perea2015sliding}
Jose~A Perea and John Harer.
\newblock Sliding windows and persistence: An application of topological methods to signal analysis.
\newblock {\em Foundations of Computational Mathematics}, 15(3):799--838, 2015.

\bibitem{maletic2016persistent}
Slobodan Maleti{\'c}, Yi~Zhao, and Milan Rajkovi{\'c}.
\newblock Persistent topological features of dynamical systems.
\newblock {\em Chaos: An Interdisciplinary Journal of Nonlinear Science}, 26(5):053105, 2016.

\bibitem{sanderson2017computational}
Nicole Sanderson, Elliott Shugerman, Samantha Molnar, James~D Meiss, and Elizabeth Bradley.
\newblock Computational topology techniques for characterizing time-series data.
\newblock In {\em International symposium on intelligent data analysis}, pages 284--296. Springer, 2017.

\bibitem{myers2019persistent}
Audun Myers, Elizabeth Munch, and Firas~A Khasawneh.
\newblock Persistent homology of complex networks for dynamic state detection.
\newblock {\em Physical Review E}, 100(2):022314, 2019.

\bibitem{mittal2017topological}
Khushboo Mittal and Shalabh Gupta.
\newblock Topological characterization and early detection of bifurcations and chaos in complex systems using persistent homology.
\newblock {\em Chaos: An Interdisciplinary Journal of Nonlinear Science}, 27(5):051102, 2017.

\bibitem{pavlova2019effects}
Olga~N Pavlova, Arkady~S Abdurashitov, Mariya~V Ulanova, Nataliya~A Shushunova, and Alexey~N Pavlov.
\newblock Effects of missing data on characterization of complex dynamics from time series.
\newblock {\em Communications in Nonlinear Science and Numerical Simulation}, 66:31--40, 2019.

\bibitem{sauer1998spurious}
Timothy~D Sauer, Joshua~A Tempkin, and James~A Yorke.
\newblock Spurious lyapunov exponents in attractor reconstruction.
\newblock {\em Physical review letters}, 81(20):4341, 1998.

\bibitem{rosenstein1993practical}
Michael~T Rosenstein, James~J Collins, and Carlo~J De~Luca.
\newblock A practical method for calculating largest lyapunov exponents from small data sets.
\newblock {\em Physica D: Nonlinear Phenomena}, 65(1-2):117--134, 1993.

\bibitem{wolf1985determining}
Alan Wolf, Jack~B Swift, Harry~L Swinney, and John~A Vastano.
\newblock Determining lyapunov exponents from a time series.
\newblock {\em Physica D: nonlinear phenomena}, 16(3):285--317, 1985.

\bibitem{ghrist2008barcodes}
Robert Ghrist.
\newblock Barcodes: the persistent topology of data.
\newblock {\em Bulletin of the American Mathematical Society}, 45(1):61--75, 2008.

\bibitem{nicolau2011topology}
Monica Nicolau, Arnold~J Levine, and Gunnar Carlsson.
\newblock Topology based data analysis identifies a subgroup of breast cancers with a unique mutational profile and excellent survival.
\newblock {\em Proceedings of the National Academy of Sciences}, 108(17):7265--7270, 2011.

\bibitem{crawford2016topological}
Lorin Crawford, Anthea Monod, Andrew~X Chen, Sayan Mukherjee, and Ra{\'u}l Rabad{\'a}n.
\newblock Topological summaries of tumor images improve prediction of disease free survival in glioblastoma multiforme.
\newblock {\em arXiv preprint arXiv:1611.06818}, 2, 2016.

\bibitem{emmett2014parametric}
Kevin Emmett, Daniel Rosenbloom, Pablo Camara, and Raul Rabadan.
\newblock Parametric inference using persistence diagrams: A case study in population genetics.
\newblock {\em arXiv preprint arXiv:1406.4582}, 2014.

\bibitem{bhattacharya2015persistent}
Subhrajit Bhattacharya, Robert Ghrist, and Vijay Kumar.
\newblock Persistent homology for path planning in uncertain environments.
\newblock {\em IEEE Transactions on Robotics}, 31(3):578--590, 2015.

\bibitem{vasudevan2013persistent}
Ramanarayan Vasudevan, Aaron Ames, and Ruzena Bajcsy.
\newblock Persistent homology for automatic determination of human-data based cost of bipedal walking.
\newblock {\em Nonlinear Analysis: Hybrid Systems}, 7(1):101--115, 2013.

\bibitem{zomorodian2001computing}
Afra~Joze Zomorodian.
\newblock {\em Computing and comprehending topology: Persistence and hierarchical morse complexes}.
\newblock University of Illinois at Urbana-Champaign, 2001.

\bibitem{vrcechmathematica}
Henry Adams and Jan Segert.
\newblock Mathematica demo on \v{C}ech and vietoris--rips complexes.

\bibitem{gudhi:RipsComplex}
Cl{\'{e}}ment Maria, Pawel Dlotko, Vincent Rouvreau, and Marc Glisse.
\newblock Rips complex.
\newblock In {\em GUDHI User and Reference Manual}. GUDHI Editorial Board, 3.9.0 edition, 2023.

\bibitem{scikittda2019}
Nathaniel Saul and Chris Tralie.
\newblock Scikit-tda: Topological data analysis for python, 2019.

\bibitem{friedman2001elements}
Jerome Friedman, Trevor Hastie, Robert Tibshirani, et~al.
\newblock {\em The elements of statistical learning}, volume~1.
\newblock Springer series in statistics New York, 2001.

\bibitem{agresti2018introduction}
Alan Agresti.
\newblock {\em An introduction to categorical data analysis}.
\newblock John Wiley \& Sons, 2018.

\bibitem{multiclass}
Peter Karsmakers, Kristiaan Pelckmans, and Johan Suykens.
\newblock Multi-class kernel logistic regression: A fixed size implementation.
\newblock {\em Proceedings of the 20th International Joint Conference on Neural Networks: 12-17 August; Orlando}, pages 1756--1761, 08 2007.

\bibitem{article}
M~Kologlu, D~Elker, Hasan Altun, and I~Sayek.
\newblock Validation of mpi and pia ii in two different groups of patients with secondary peritonitis.
\newblock {\em Hepato-gastroenterology}, 48:147--51, 11 2000.

\bibitem{BIONDO2000635}
Sebastiano Biondo, Emilio Ramos, Manuel Deiros, Juan, Javier, Pablo Moreno, Leandre Farran, and Eduardo Jaurrieta.
\newblock Prognostic factors for mortality in left colonic peritonitis: a new scoring system11no competing interests declared.
\newblock {\em Journal of the American College of Surgeons}, 191(6):635--642, 2000.

\bibitem{truett1967multivariate}
Jeanne Truett, Jerome Cornfield, and William Kannel.
\newblock A multivariate analysis of the risk of coronary heart disease in framingham.
\newblock {\em Journal of chronic diseases}, 20(7):511--524, 1967.

\bibitem{harrell2017regression}
Frank~E Harrell.
\newblock {\em Regression modeling strategies}, volume 330.
\newblock Springer, 2017.

\bibitem{PALEI200988}
Sanjay~Kumar Palei and Samir~Kumar Das.
\newblock Logistic regression model for prediction of roof fall risks in bord and pillar workings in coal mines: An approach.
\newblock {\em Safety Science}, 47(1):88--96, 2009.

\bibitem{raschka2019python}
Sebastian Raschka and Vahid Mirjalili.
\newblock {\em Python machine learning: Machine learning and deep learning with Python, scikit-learn, and TensorFlow 2}.
\newblock Packt Publishing Ltd, 2019.

\bibitem{sklearn_api}
Lars Buitinck, Gilles Louppe, Mathieu Blondel, Fabian Pedregosa, Andreas Mueller, Olivier Grisel, Vlad Niculae, Peter Prettenhofer, Alexandre Gramfort, Jaques Grobler, Robert Layton, Jake VanderPlas, Arnaud Joly, Brian Holt, and Ga{\"{e}}l Varoquaux.
\newblock {API} design for machine learning software: experiences from the scikit-learn project.
\newblock In {\em ECML PKDD Workshop: Languages for Data Mining and Machine Learning}, pages 108--122, 2013.

\bibitem{jaquette2020fractal}
Jonathan Jaquette and Benjamin Schweinhart.
\newblock Fractal dimension estimation with persistent homology: a comparative study.
\newblock {\em Communications in Nonlinear Science and Numerical Simulation}, 84:105163, 2020.

\bibitem{jiang2022topological}
Fangfang Jiang, Bowen Xu, Ziyu Zhu, and Biyong Zhang.
\newblock Topological data analysis approach to extract the persistent homology features of ballistocardiogram signal in unobstructive atrial fibrillation detection.
\newblock {\em IEEE Sensors Journal}, 22(7):6920--6930, 2022.

\bibitem{david2000applied}
David~W.. Hosmer, Stanley Lemeshow, and Rodney~X.. Sturdivant.
\newblock {\em Applied logistic regression}.
\newblock Wiley New York, 2000.

\bibitem{madow1944theory}
William~G Madow and Lillian~H Madow.
\newblock On the theory of systematic sampling, i.
\newblock {\em The Annals of Mathematical Statistics}, 15(1):1--24, 1944.

\bibitem{edelsbrunner2000topological}
Herbert Edelsbrunner, David Letscher, and Afra Zomorodian.
\newblock Topological persistence and simplification.
\newblock In {\em Proceedings 41st annual symposium on foundations of computer science}, pages 454--463. IEEE, 2000.

\bibitem{zomorodian2005topology}
Afra~J Zomorodian.
\newblock {\em Topology for computing}, volume~16.
\newblock Cambridge university press, 2005.

\bibitem{delfinado1995incremental}
Cecil Jose~A Delfinado and Herbert Edelsbrunner.
\newblock An incremental algorithm for betti numbers of simplicial complexes on the 3-sphere.
\newblock {\em Computer Aided Geometric Design}, 12(7):771--784, 1995.

\bibitem{kramar2016analysis}
Miroslav Kram{\'a}r, Rachel Levanger, Jeffrey Tithof, Balachandra Suri, Mu~Xu, Mark Paul, Michael~F Schatz, and Konstantin Mischaikow.
\newblock Analysis of kolmogorov flow and rayleigh--b{\'e}nard convection using persistent homology.
\newblock {\em Physica D: Nonlinear Phenomena}, 334:82--98, 2016.

\bibitem{chan2013topology}
Joseph~Minhow Chan, Gunnar Carlsson, and Raul Rabadan.
\newblock Topology of viral evolution.
\newblock {\em Proceedings of the National Academy of Sciences}, 110(46):18566--18571, 2013.

\bibitem{xia2014persistent}
Kelin Xia and Guo-Wei Wei.
\newblock Persistent homology analysis of protein structure, flexibility, and folding.
\newblock {\em International journal for numerical methods in biomedical engineering}, 30(8):814--844, 2014.

\bibitem{bendich2016persistent}
Paul Bendich, James~S Marron, Ezra Miller, Alex Pieloch, and Sean Skwerer.
\newblock Persistent homology analysis of brain artery trees.
\newblock {\em The annals of applied statistics}, 10(1):198, 2016.

\bibitem{tralie2018ripser}
Christopher Tralie, Nathaniel Saul, and Rann Bar-On.
\newblock Ripser. py: A lean persistent homology library for python.
\newblock {\em Journal of Open Source Software}, 3(29):925, 2018.

\bibitem{maria2014gudhi}
Clement Maria, Jean-Daniel Boissonnat, Marc Glisse, and Mariette Yvinec.
\newblock The gudhi library: Simplicial complexes and persistent homology.
\newblock In {\em International congress on mathematical software}, pages 167--174. Springer, 2014.

\bibitem{atienza2019persistent}
Maria~Nieves Atienza~Martinez, Rocio Gonzalez~Diaz, and Matteo Rucco.
\newblock Persistent entropy for separating topological features from noise in vietoris-rips complexes.
\newblock {\em Journal of Intelligent Information Systems, 52 (3), 637-655.}, 2019.

\bibitem{chintakunta2015entropy}
Harish Chintakunta, Thanos Gentimis, Rocio Gonzalez-Diaz, Maria-Jose Jimenez, and Hamid Krim.
\newblock An entropy-based persistence barcode.
\newblock {\em Pattern Recognition}, 48(2):391--401, 2015.

\bibitem{gottwald2004new}
Georg~A Gottwald and Ian Melbourne.
\newblock A new test for chaos in deterministic systems.
\newblock {\em Proceedings of the Royal Society of London. Series A: Mathematical, Physical and Engineering Sciences}, 460(2042):603--611, 2004.

\bibitem{fouda2013three}
JS~Armand~Eyebe Fouda, J~Yves Effa, Martin Kom, and Maaruf Ali.
\newblock The three-state test for chaos detection in discrete maps.
\newblock {\em Applied Soft Computing}, 13(12):4731--4737, 2013.

\bibitem{tang2004geometric}
Liying Tang and Mark Crovella.
\newblock Geometric exploration of the landmark selection problem.
\newblock In {\em International Workshop on Passive and Active Network Measurement}, pages 63--72. Springer, 2004.

\bibitem{carlsson2008local}
Gunnar Carlsson, Tigran Ishkhanov, Vin De~Silva, and Afra Zomorodian.
\newblock On the local behavior of spaces of natural images.
\newblock {\em International journal of computer vision}, 76(1):1--12, 2008.

\bibitem{comaniciu2002mean}
Dorin Comaniciu and Peter Meer.
\newblock Mean shift: A robust approach toward feature space analysis.
\newblock {\em IEEE Transactions on pattern analysis and machine intelligence}, 24(5):603--619, 2002.

\bibitem{scikit-learn}
F.~Pedregosa, G.~Varoquaux, A.~Gramfort, V.~Michel, B.~Thirion, O.~Grisel, M.~Blondel, P.~Prettenhofer, R.~Weiss, V.~Dubourg, J.~Vanderplas, A.~Passos, D.~Cournapeau, M.~Brucher, M.~Perrot, and E.~Duchesnay.
\newblock Scikit-learn: Machine learning in {P}ython.
\newblock {\em Journal of Machine Learning Research}, 12:2825--2830, 2011.

\bibitem{kovacev2016using}
Violeta Kovacev-Nikolic, Peter Bubenik, Dragan Nikoli{\'c}, and Giseon Heo.
\newblock Using persistent homology and dynamical distances to analyze protein binding.
\newblock {\em Statistical applications in genetics and molecular biology}, 15(1):19--38, 2016.

\bibitem{gameiro2015topological}
Marcio Gameiro, Yasuaki Hiraoka, Shunsuke Izumi, Miroslav Kramar, Konstantin Mischaikow, and Vidit Nanda.
\newblock A topological measurement of protein compressibility.
\newblock {\em Japan Journal of Industrial and Applied Mathematics}, 32:1--17, 2015.

\bibitem{vietoris1927hoheren}
Leopold Vietoris.
\newblock {\"U}ber den h{\"o}heren zusammenhang kompakter r{\"a}ume und eine klasse von zusammenhangstreuen abbildungen.
\newblock {\em Mathematische Annalen}, 97(1):454--472, 1927.

\bibitem{wolpert1997no}
David~H Wolpert and William~G Macready.
\newblock No free lunch theorems for optimization.
\newblock {\em IEEE transactions on evolutionary computation}, 1(1):67--82, 1997.

\end{thebibliography}

\end{document}